\documentclass[fleqn,usenatbib]{mnras}

\usepackage{newtxtext,newtxmath}
\usepackage{amsmath}
\usepackage{graphicx}
\usepackage[T1]{fontenc}
\DeclareRobustCommand{\VAN}[3]{#2}
\let\VANthebibliography\thebibliography
\def\thebibliography{\DeclareRobustCommand{\VAN}[3]{##3}\VANthebibliography}

\title[Updating the Ephemeris and Physical Properties of Five Long-period Transiting Exoplanets]{Updating the Ephemeris and Physical Properties of Five Long-period Transiting Exoplanets Using TESS and CHEOPS}

\author[Suman Saha]{
Suman Saha,$^{1}$\thanks{Email: suman.saha@mail.udp.cl}
\\
$^{1}$Instituto de Estudios Astrofísicos, Facultad de Ingeniería Ciencias, Universidad Diego Portales, Av. Ejército Libertador 441, Santiago, Chile
}

\begin{document}
\label{firstpage}
\pagerange{\pageref{firstpage}--\pageref{lastpage}}
\maketitle

\begin{abstract}
The transiting long-period exoplanets are the most interesting follow-up targets for the next-generation instruments, using the most sophisticated observational techniques. However, the scarcity in their transit events often leads to larger uncertainties in their known ephemeris, also resulting in larger biases in their known physical properties in many cases. In this work, I have used the newer publicly available observations from TESS and CHEOPS for five very interesting long-period transiting exoplanets, i.e., HD95338 b, TOI-2134 c, K2-290 c, TOI-1898 b, and TOI-813 b, combined with the previously reported observations, to reanalyze their transit properties and estimate the updated ephemeris. The analyses also incorporated a critical noise treatment algorithm, which uses well-tested techniques such as wavelet denoising and Gaussian process regression, to effectively reduce the impact of various noise components present in the lightcurves on the estimated parameters. The study has resulted in a more precise estimation of ephemeris for all the targets, with the precision in the estimated periods being better than 5 seconds, except for TOI-813 b, for which the precision in the estimated period is better than 21 seconds. The other transit parameters also got updated, with statistically significant improvements seen in most of the cases, the major notable improvements being in the estimated value of the impact parameter of TOI-1898 b, the orbital semi-major axis of TOI-2134 c, and the radius of HD95338 b. Although long-period exoplanets are expected to show more significant transit timing variations in the presence of other undetected planetary mass objects, no such variations were recorded for these targets.
\end{abstract}

\begin{keywords}
Planetary systems -- Planets and satellites: gaseous planets -- Techniques: photometric
\end{keywords}

\section{Introduction}\label{sec:sec1}

The long-period exoplanets (P $>$ 40 days) are thought to be the key to answering several recurring questions in the existing theories of planet formation and evolution. Unlike the close-in exoplanets, they are not affected by heavy irradiation by the host stars and the photoevaporation of their atmospheres \cite[e.g.][]{2011A&A...532A...6S, 2019AREPS..47...67O}. Thus, they are likely to have retained their original atmospheric compositions, which could be studied to better understand the planetary origins. Several long-period exoplanets have been found in highly eccentric orbits \cite[e.g.][]{2022AJ....163...61D, 2024MNRAS.527.5385R}, which could have resulted from phenomena like planet-planet scattering \cite[e.g.][]{2008ApJ...686..580C, 2019A&A...629L...7C} and the Lidov-Kozai effect \cite[e.g.][]{2011ApJ...742...94L, 2016ARA&A..54..441N}. Unlike the close-in exoplanets, their orbits are not altered by tidal dampening \cite[e.g.][]{2010exop.book..239C}, and thus the impact and magnitude of these phenomena on planetary migration can be studied. Due to their weaker gravitational interaction with the host stars, in contrast to their close-in counter-parts, the long-period exoplanets are known to show large Transit Timing Variations (TTVs) from the presence of other planetary-mass bodies in these systems \cite[e.g.][]{2023AJ....166..201H}. This makes the long-term follow-up of these planets a valuable asset for the discovery of undetected planetary mass objects in these systems, which will also help to understand the general demographics of the exoplanetary systems in a better way. Moreover, long-period exoplanets have the potential to host exomooons, as they retain much larger Hill's spheres compared to the close-in exoplanets \cite[e.g.][]{2010ApJ...719L.145N, 2021PASP..133i4401D, 2022ApJ...936....2S}, which increases the probability of the survival of natural satellites around them. Detection of exomoons can revolutionize our understanding of planet formation and evolution mechanisms.

\begin{figure*}
	\centering
    \begin{tabular}{cc}
	\includegraphics[width=0.48\linewidth]{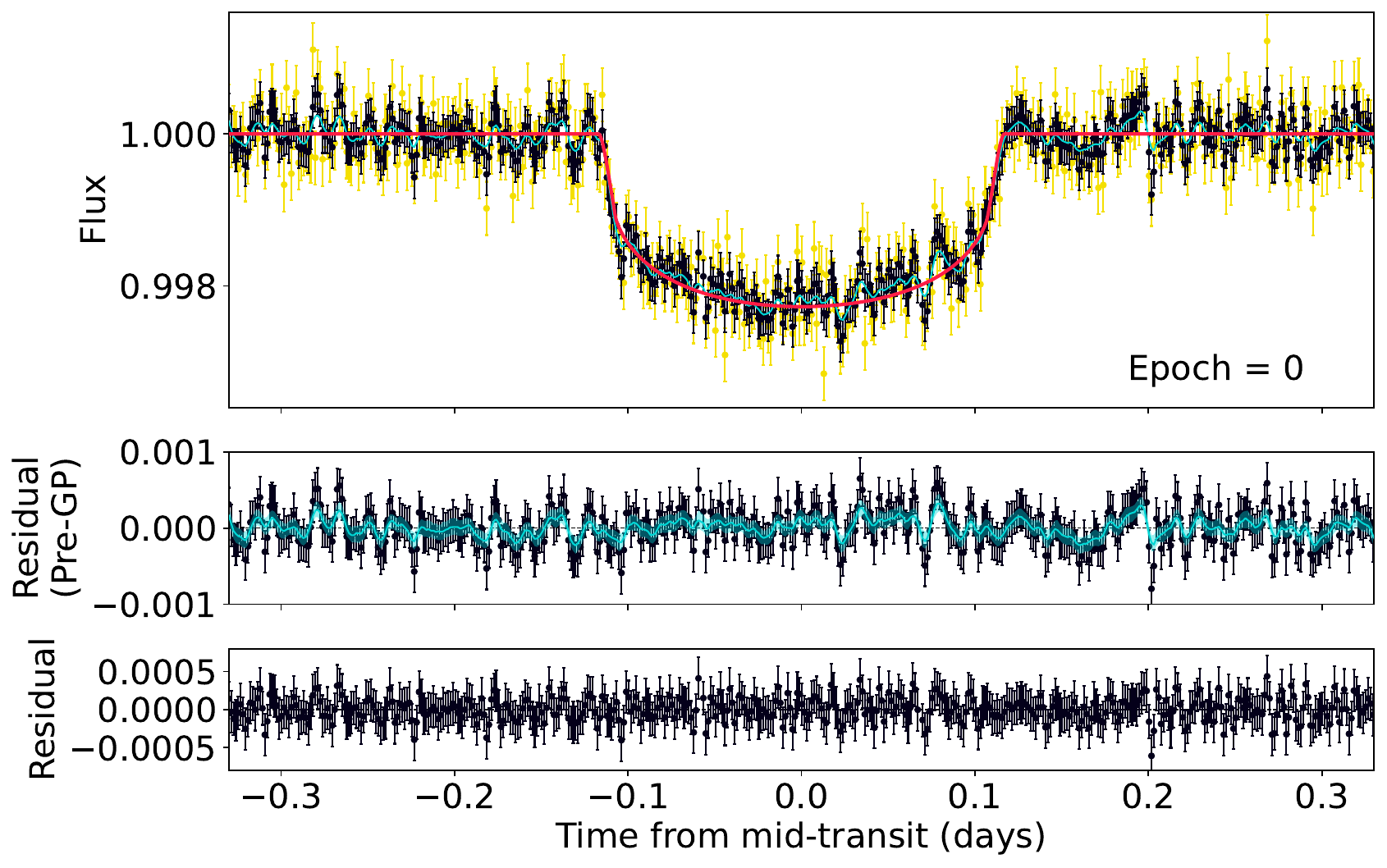}\vspace{0.3cm} &
    \includegraphics[width=0.48\linewidth]{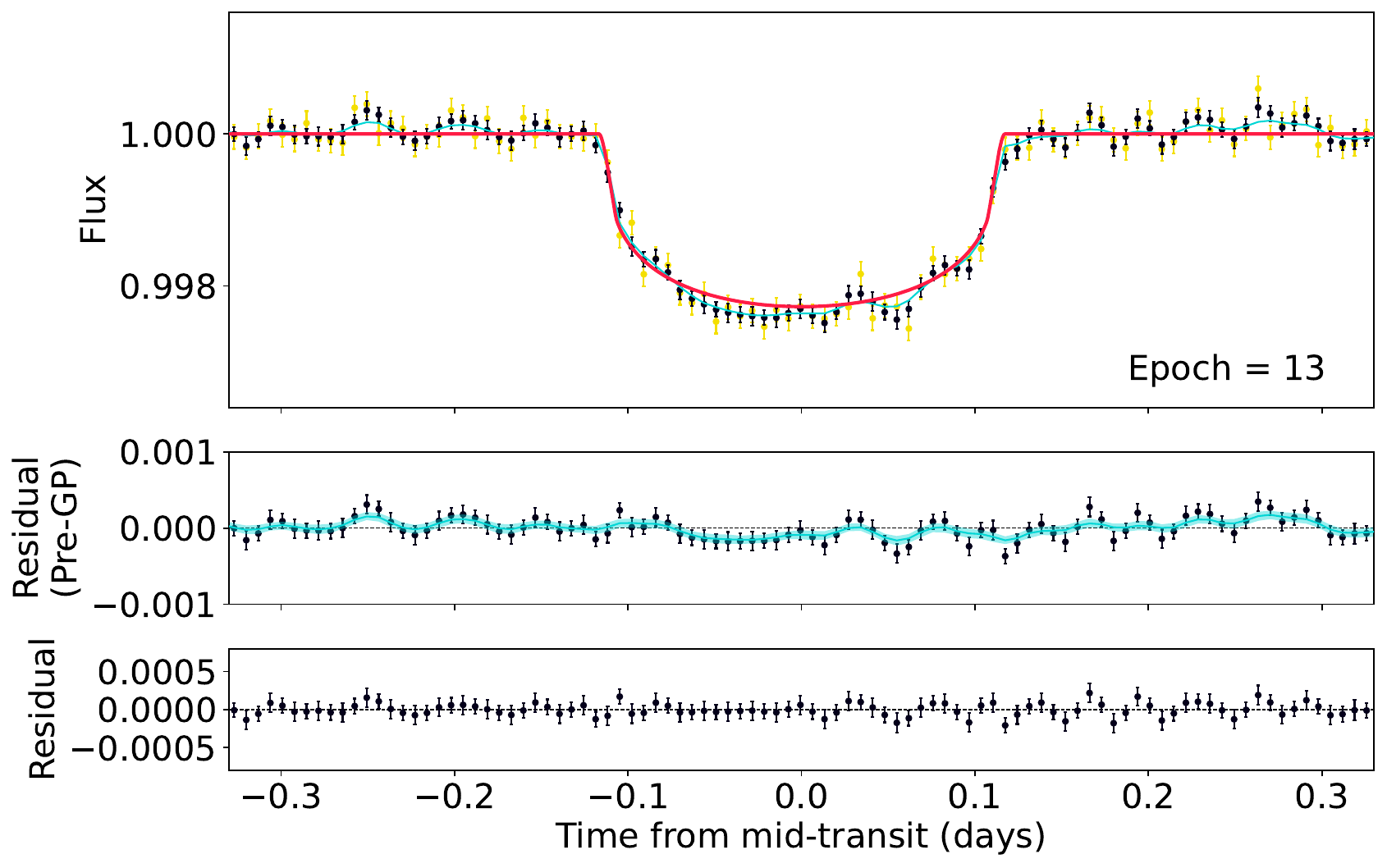}\vspace{0.3cm} \\
    \includegraphics[width=0.48\linewidth]{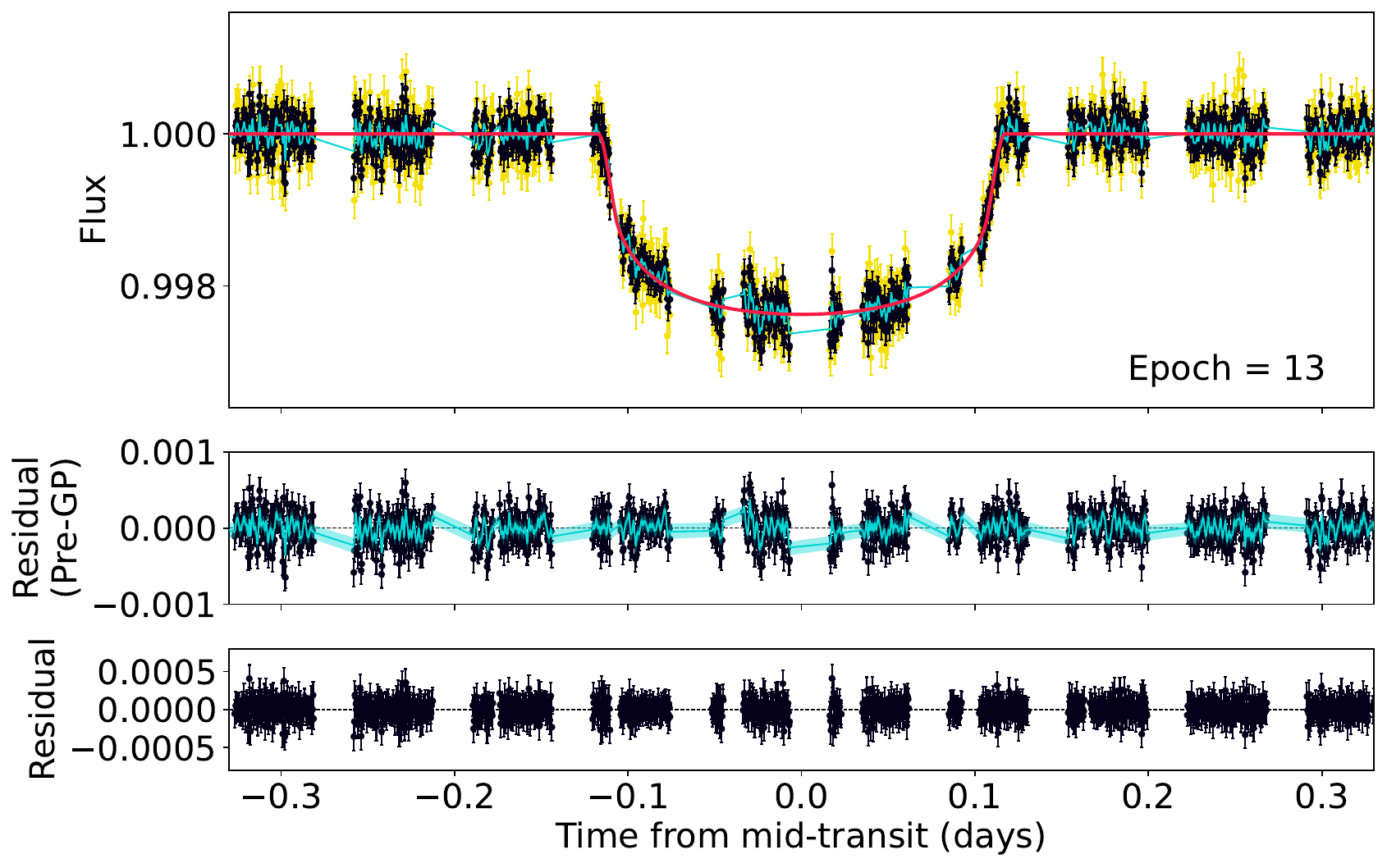}\vspace{0.3cm} &
    \includegraphics[width=0.48\linewidth]{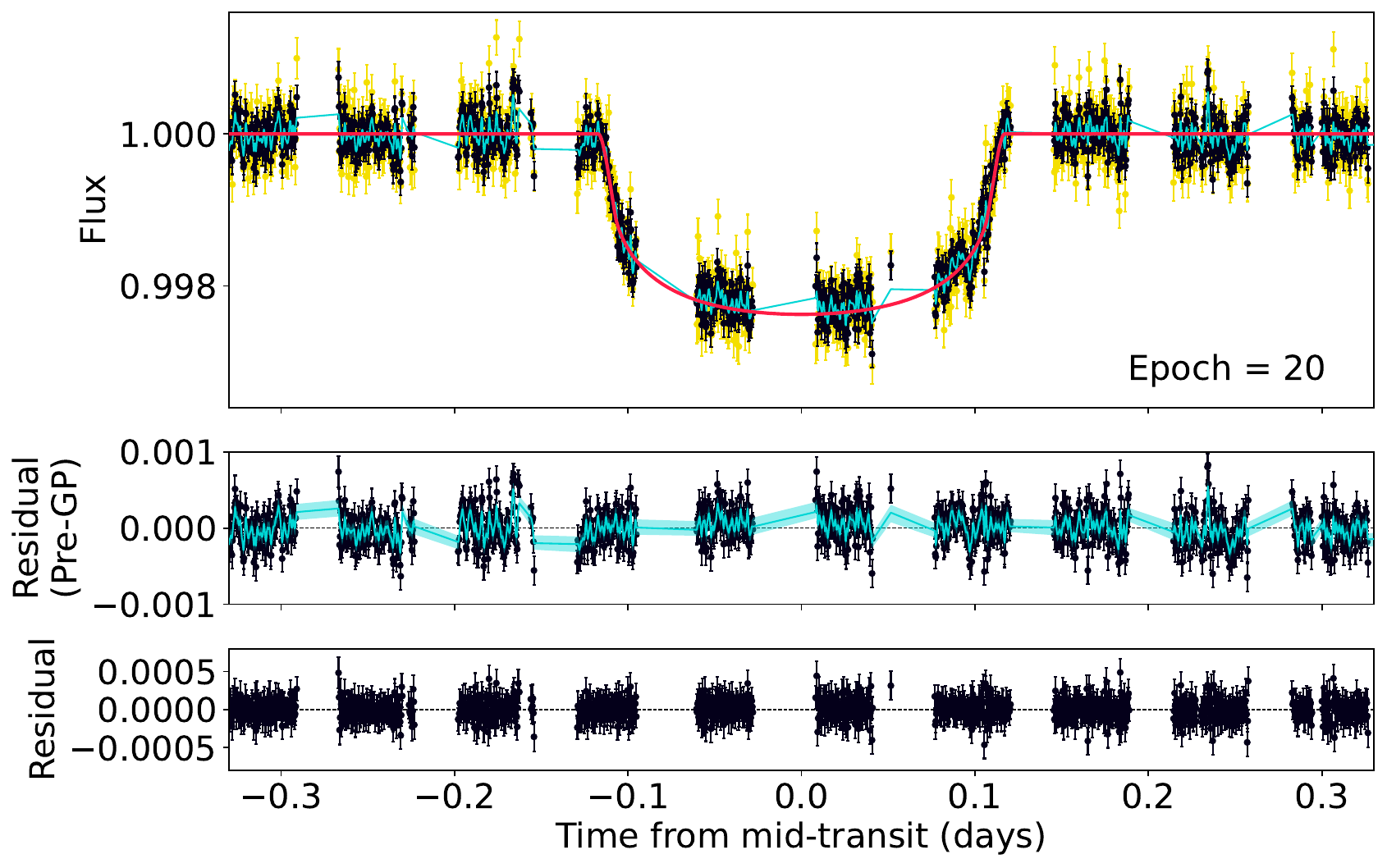}\vspace{0.3cm} \\
    \includegraphics[width=0.48\linewidth]{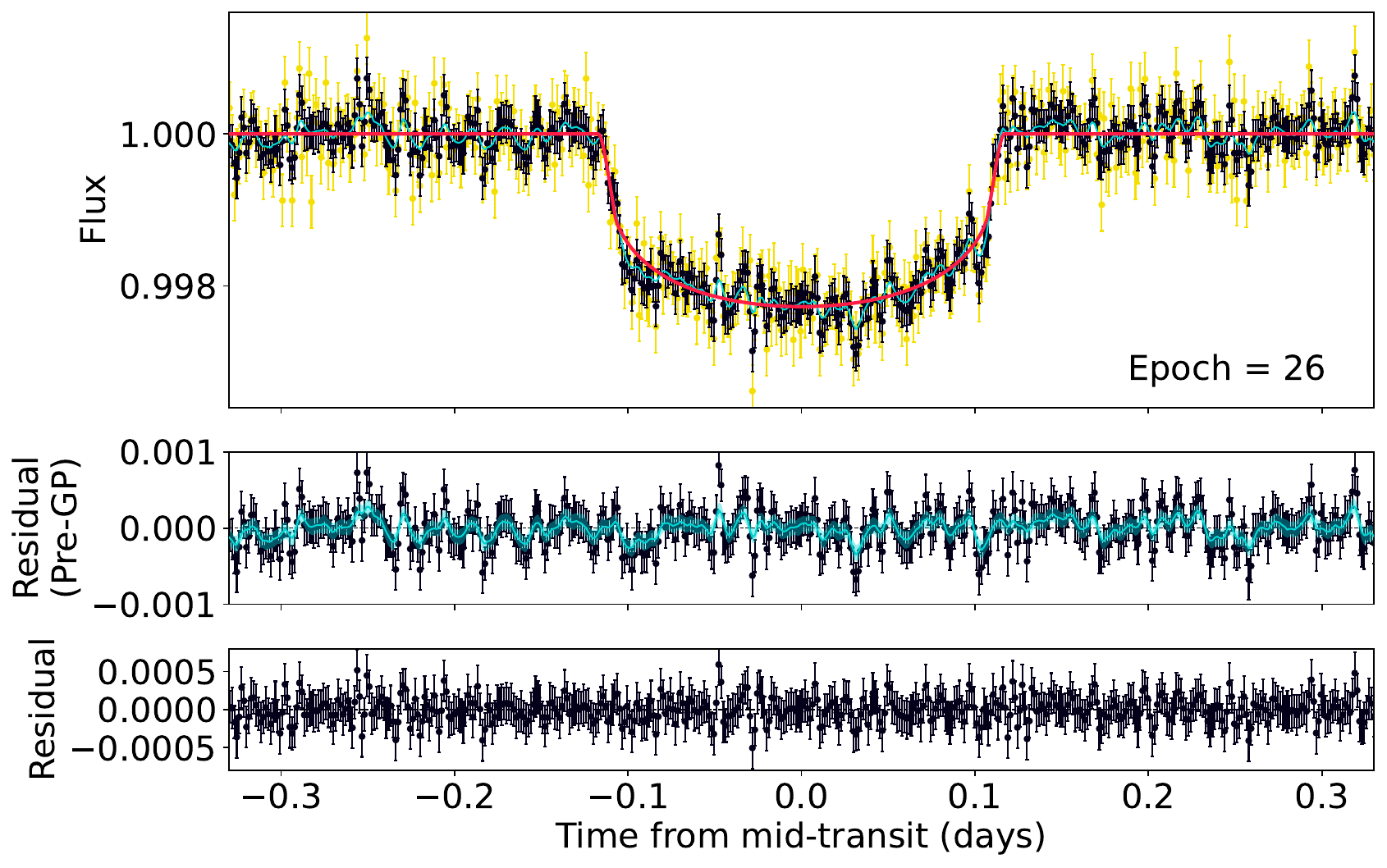}\vspace{0.2cm} &
    \includegraphics[width=0.48\linewidth]{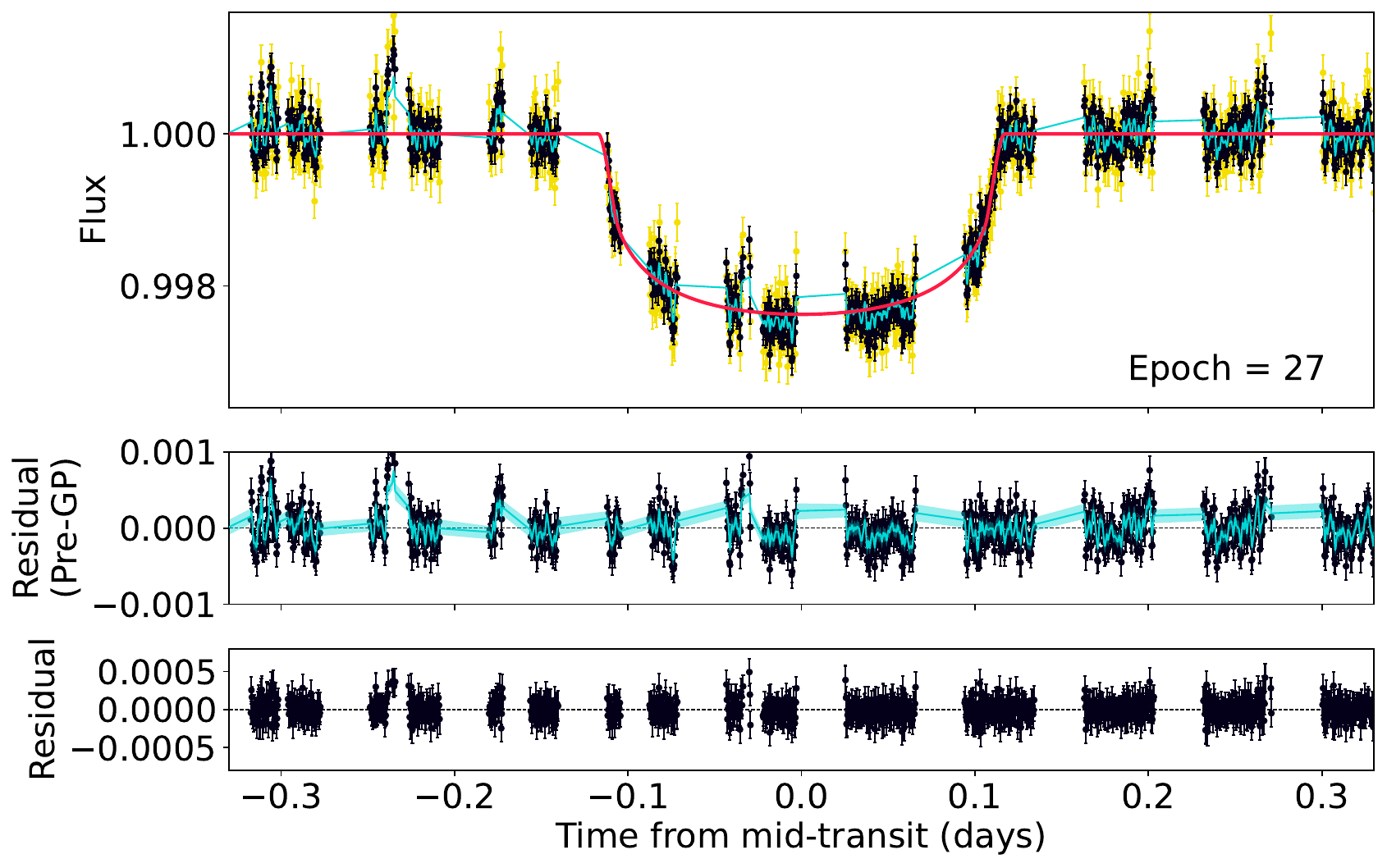}\vspace{0.2cm}
    \end{tabular}
	\caption{Observed and best-fit model light-curves for HD95338 b. For each transit, the top panel shows the unprocessed lightcurve (yellow), the lightcurve after wavelet denoising (black), the best-fit GP+transit model (cyan), and the best-fit transit model (red); the middle panel shows the mean residual flux before GP regression (black), and the mean (cyan) and 1-$\sigma$ interval (shaded cyan) of the best-fit GP regression model; and the bottom panel shows the final residual flux (black), corresponding to the best-fit GP+transit model. The epochs are defined with respect to the first detected transit of the planet.}
	\label{fig:fig1}
\end{figure*}

\begin{figure*}
	\centering
    \begin{tabular}{cc}
	\includegraphics[width=0.48\linewidth]{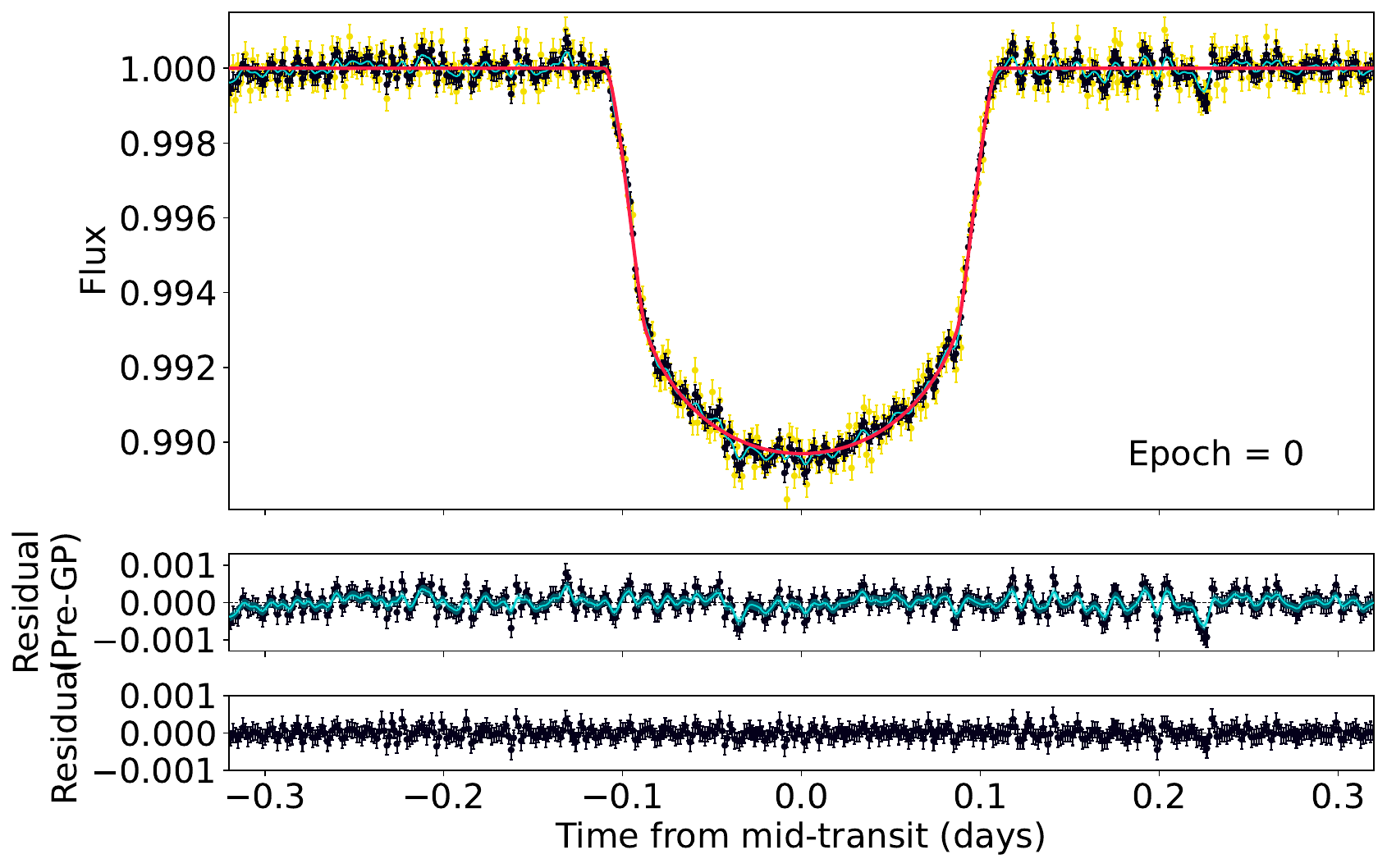}\vspace{0.3cm} &
    \includegraphics[width=0.48\linewidth]{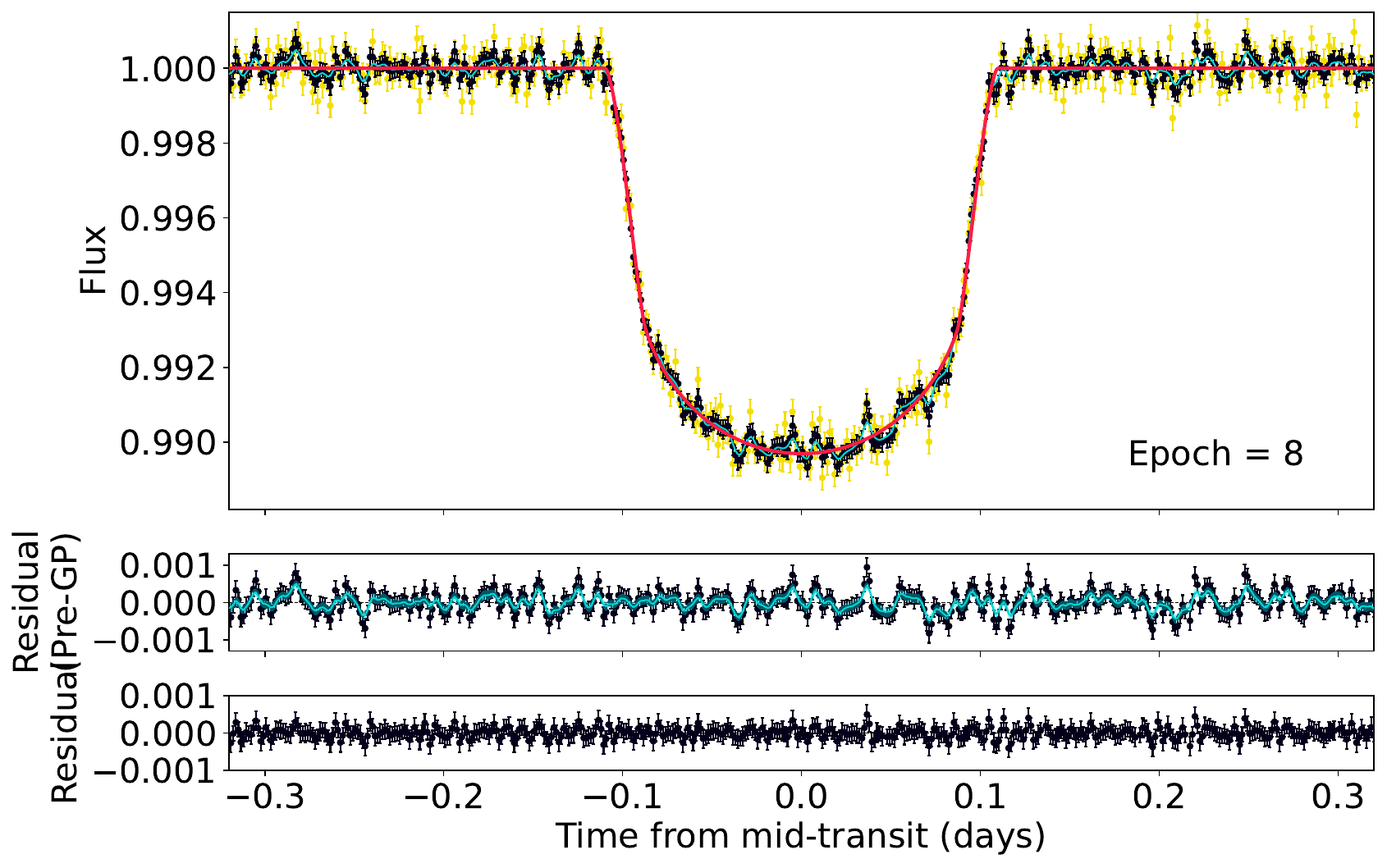}\vspace{0.3cm}
    \end{tabular}
	\caption{Same as Figure \ref{fig:fig1}, but for TOI-2134 c.}
	\label{fig:fig2}
\end{figure*}

\begin{figure*}
	\centering
    \begin{tabular}{cc}
	\includegraphics[width=0.48\linewidth]{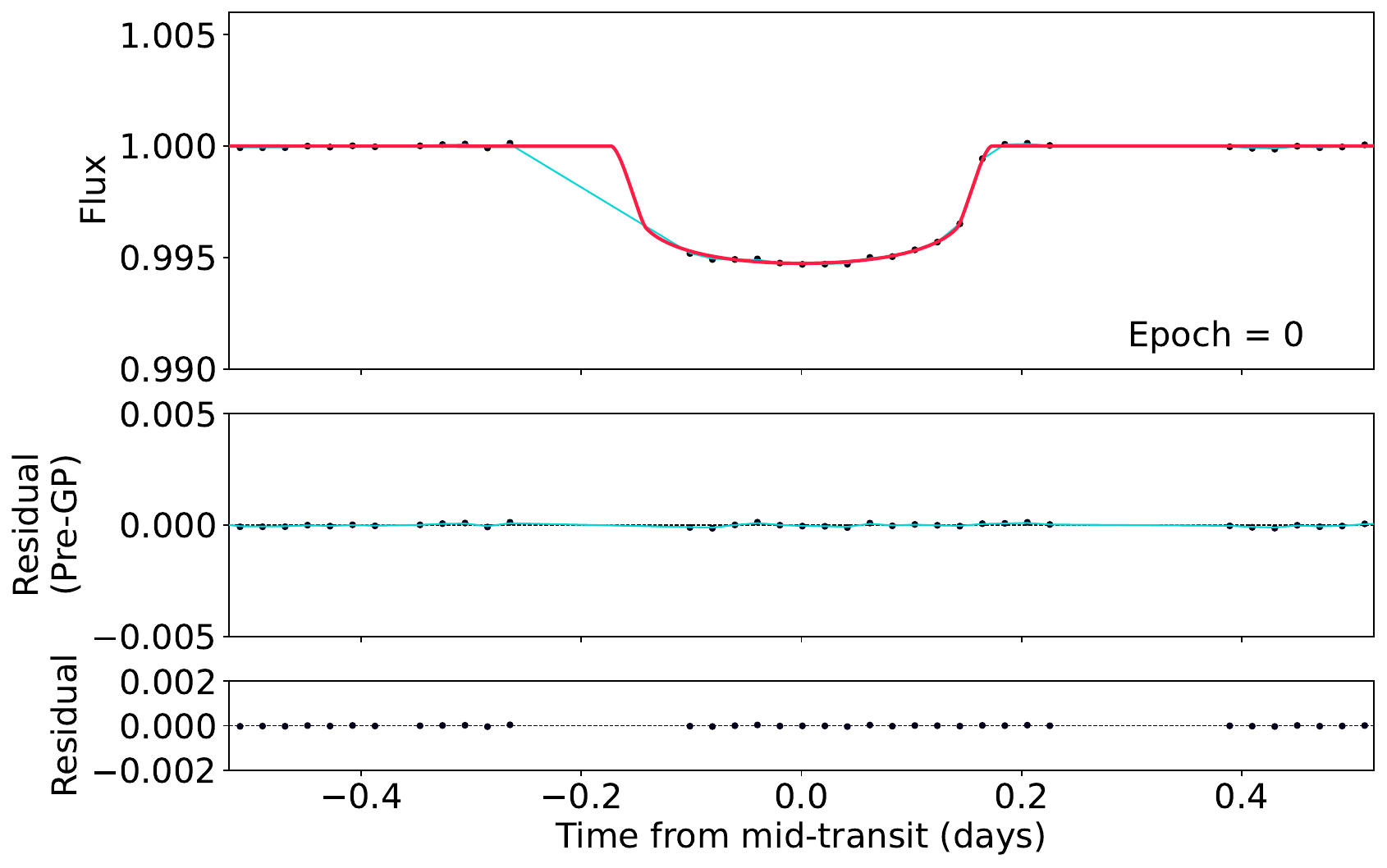}\vspace{0.3cm} &
    \includegraphics[width=0.48\linewidth]{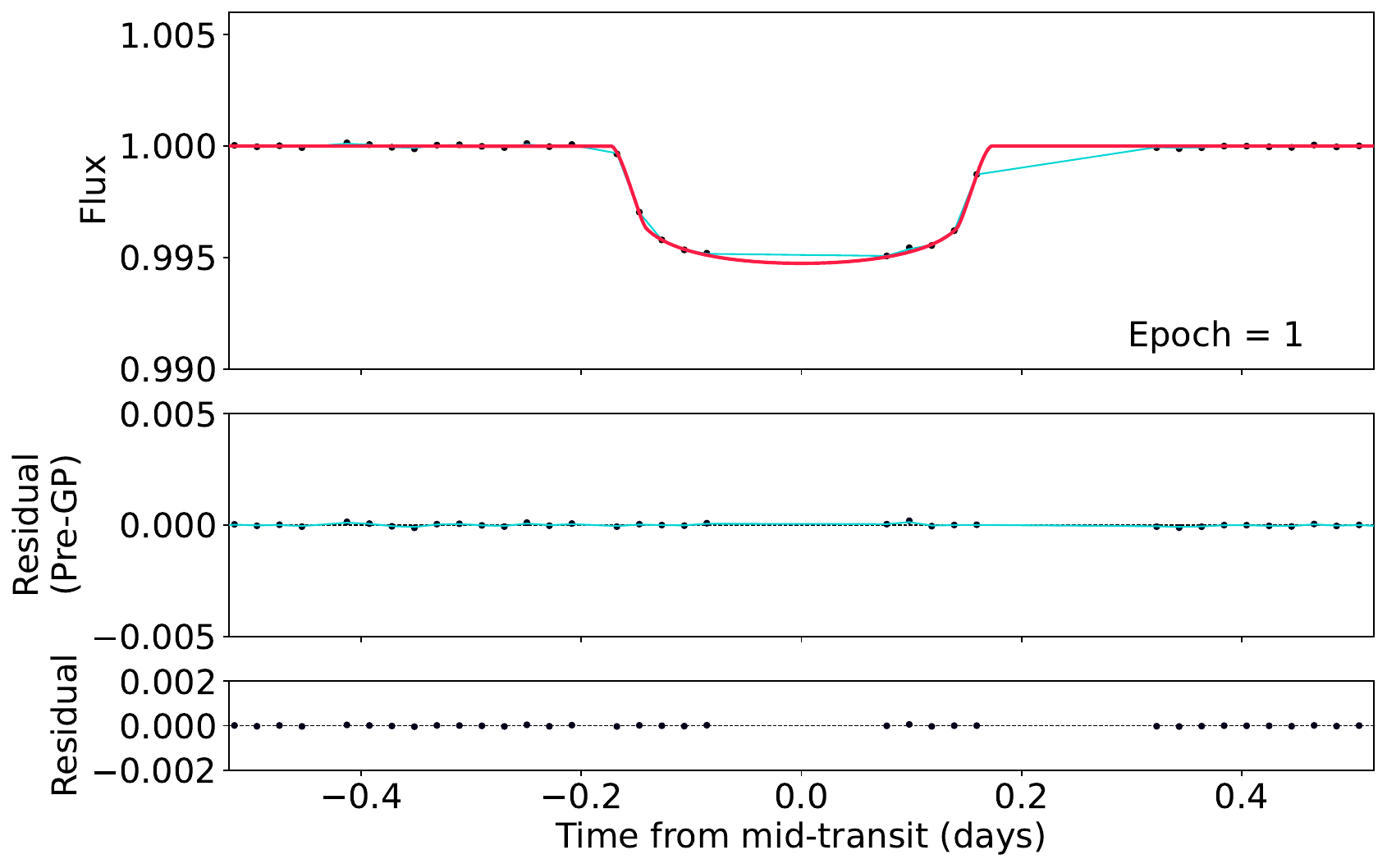}\vspace{0.3cm} \\
    \includegraphics[width=0.48\linewidth]{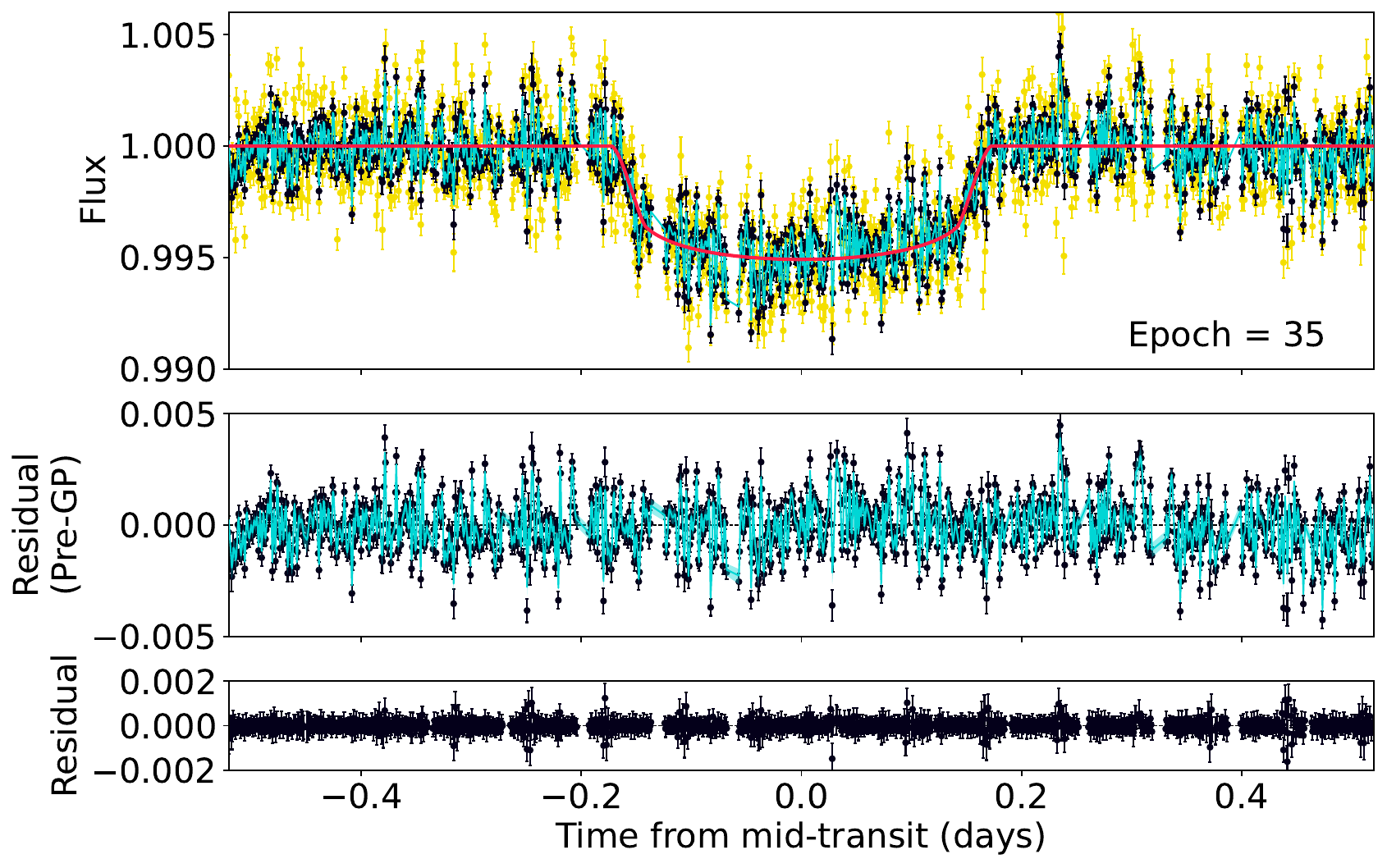}\vspace{0.2cm} &
    \includegraphics[width=0.48\linewidth]{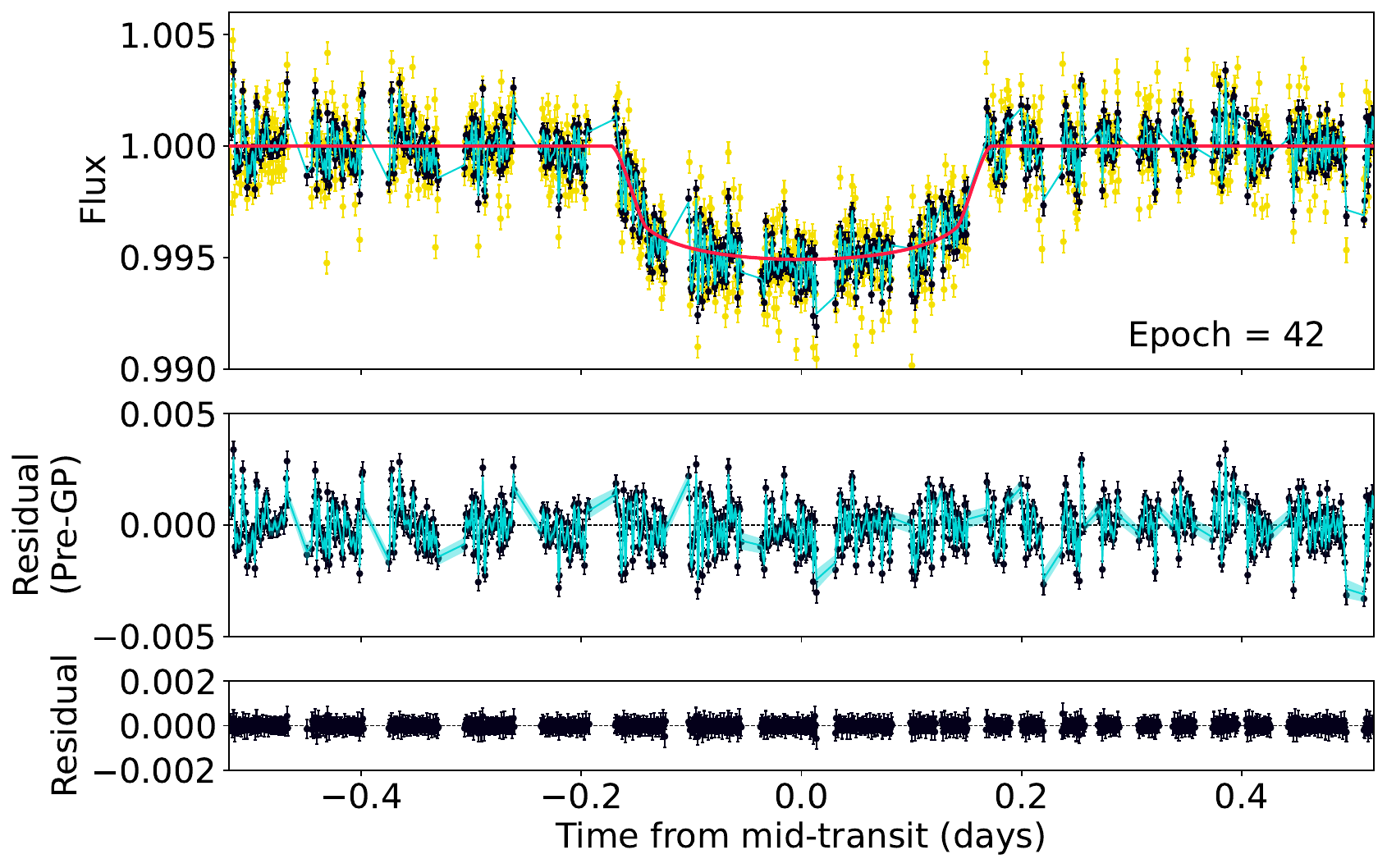}\vspace{0.2cm}
    \end{tabular}
	\caption{Same as Figure \ref{fig:fig1}, but for K2-290 c. The Y-axis scale has been kept consistent across epochs to emphasize the difference between low-cadence K2 and high-cadence CHEOPS lightcurves.}
	\label{fig:fig3}
\end{figure*}

\begin{figure*}
	\centering
    \begin{tabular}{cc}
	\includegraphics[width=0.48\linewidth]{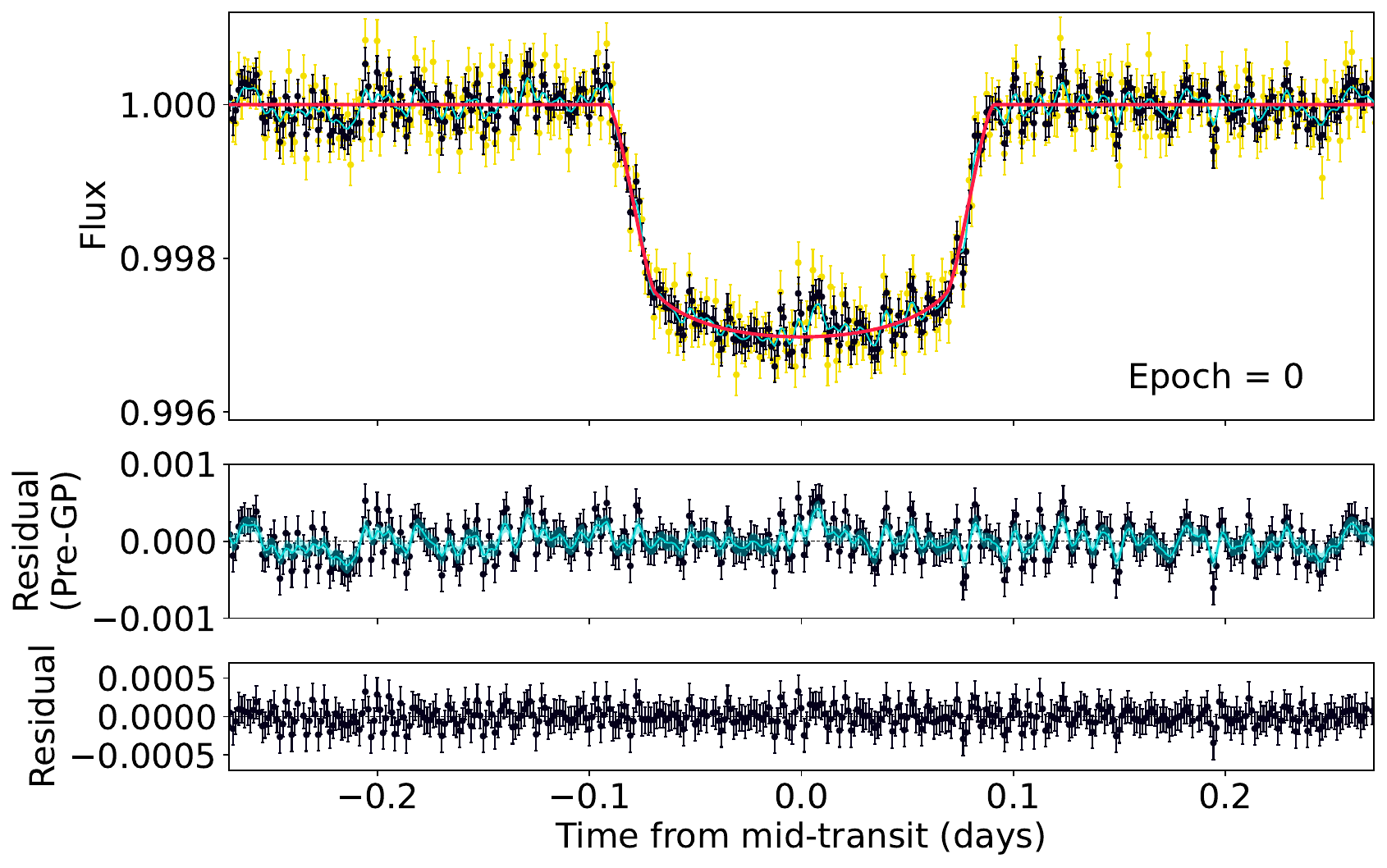}\vspace{0.3cm} &
    \includegraphics[width=0.48\linewidth]{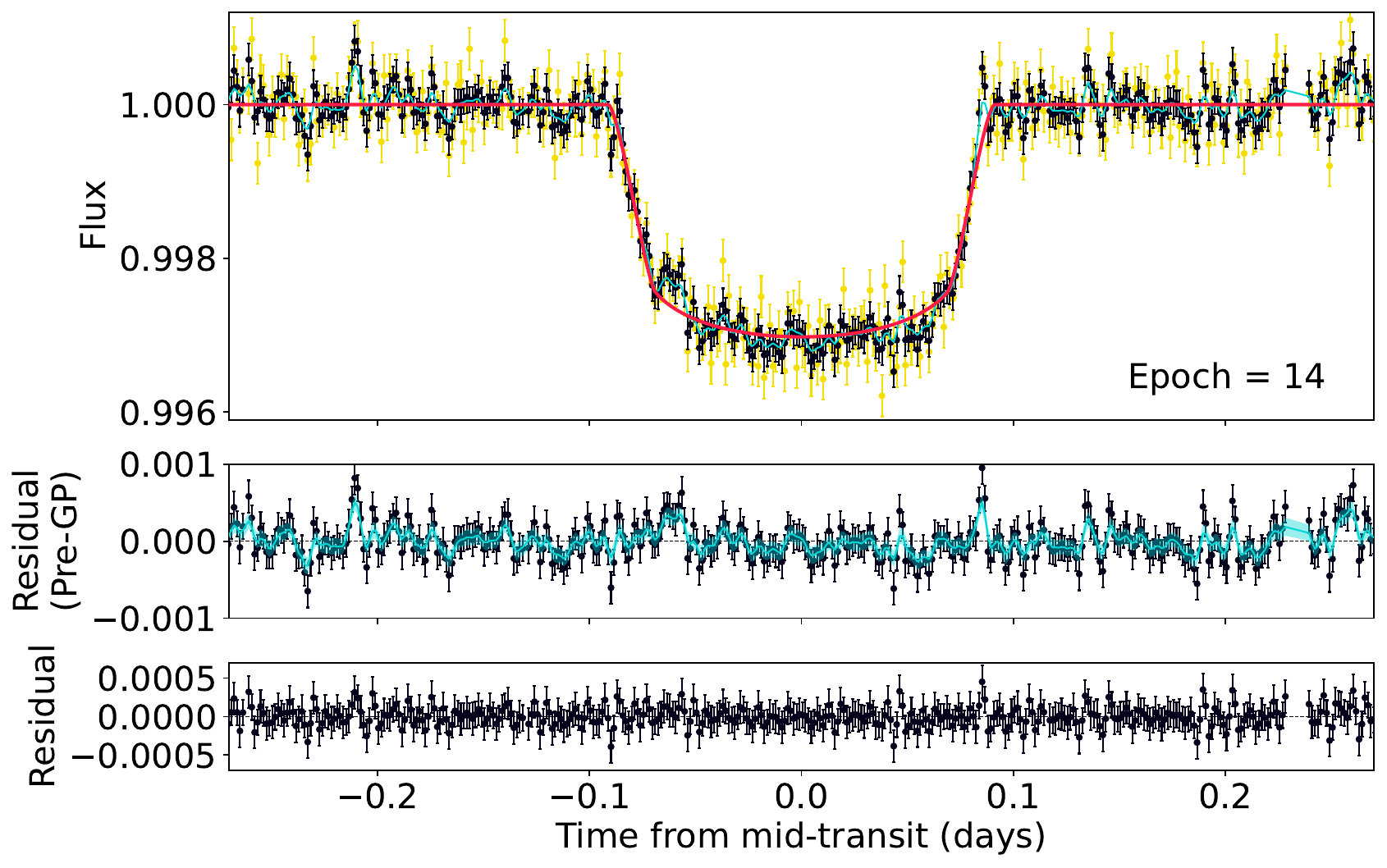}\vspace{0.3cm} \\
    \includegraphics[width=0.48\linewidth]{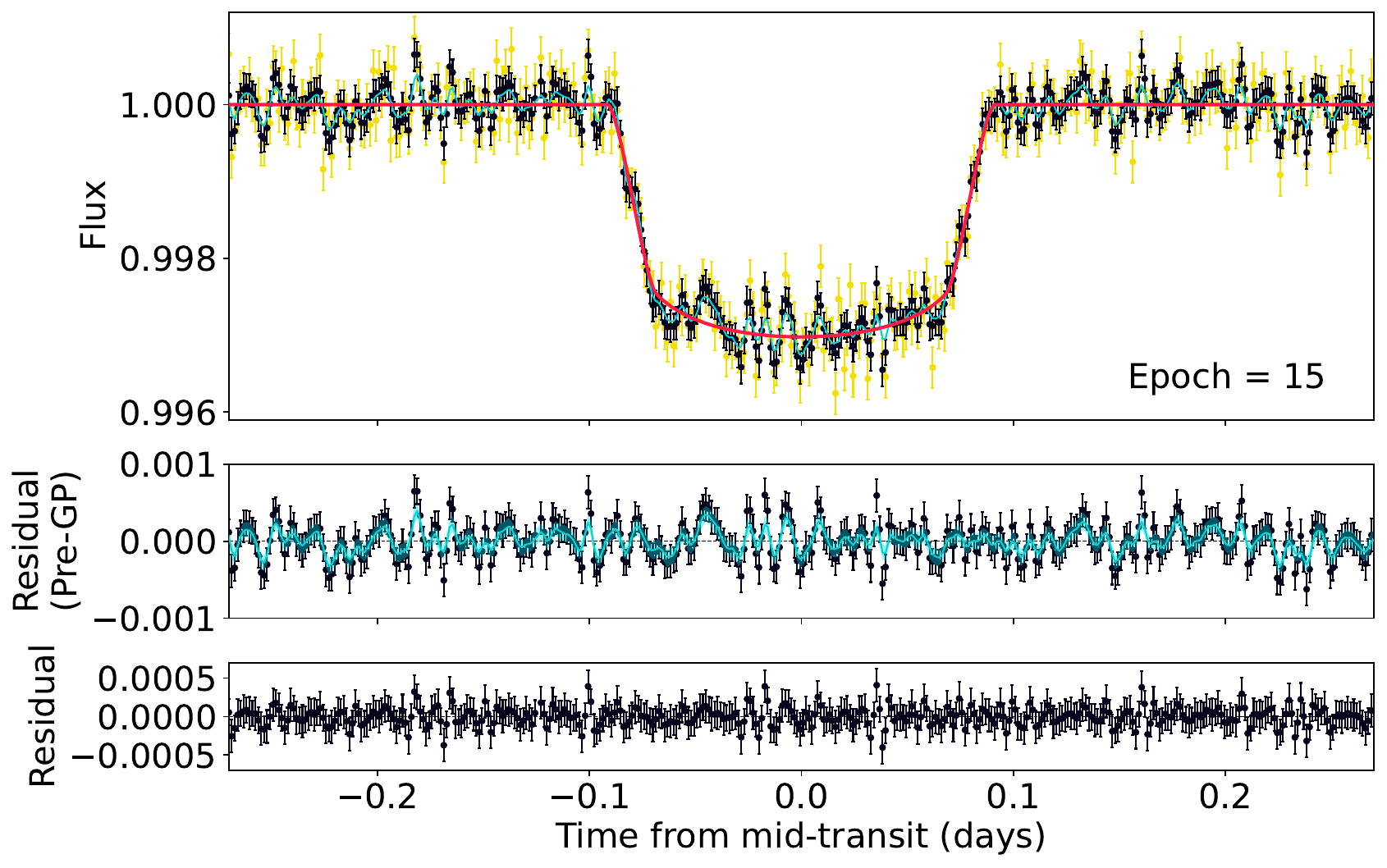}\vspace{0.3cm} &
    \includegraphics[width=0.48\linewidth]{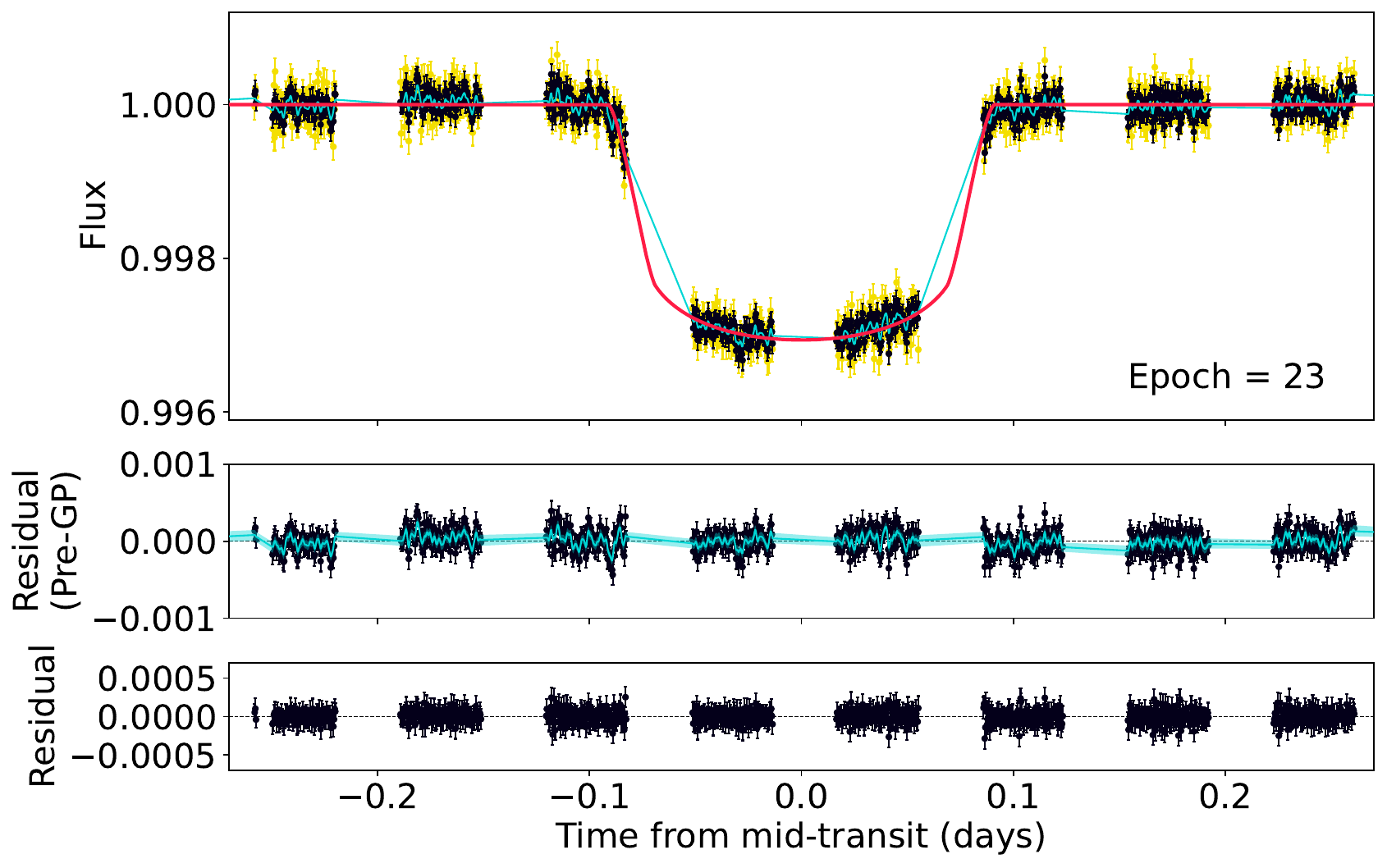}\vspace{0.3cm} \\
    \includegraphics[width=0.48\linewidth]{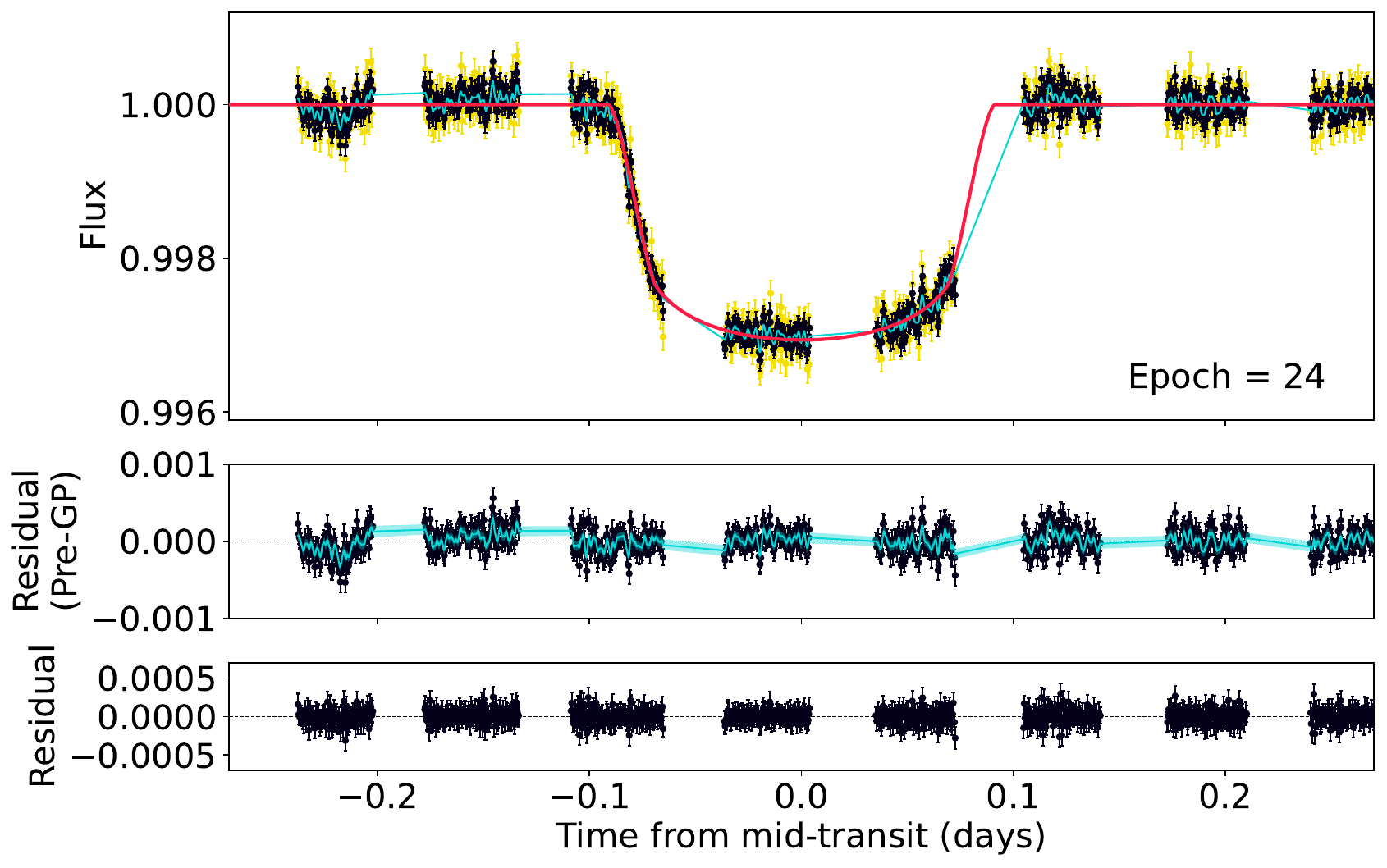}\vspace{0.2cm} &
    \includegraphics[width=0.48\linewidth]{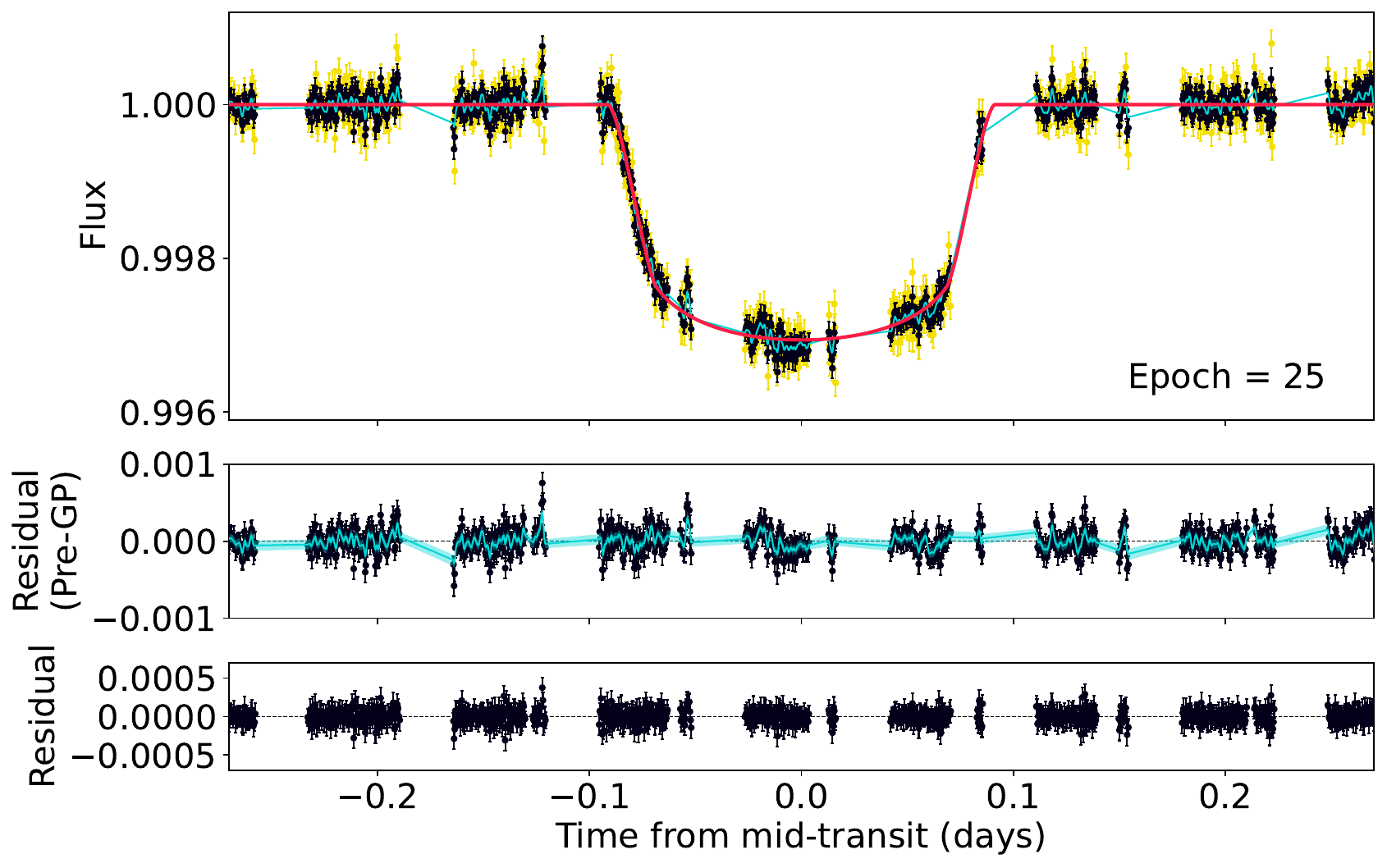}\vspace{0.2cm}
    \end{tabular}
	\caption{Same as Figure \ref{fig:fig1}, but for TOI-1898 b.}
	\label{fig:fig4}
\end{figure*}

\begin{figure*}
	\centering
    \begin{tabular}{cc}
	\includegraphics[width=0.48\linewidth]{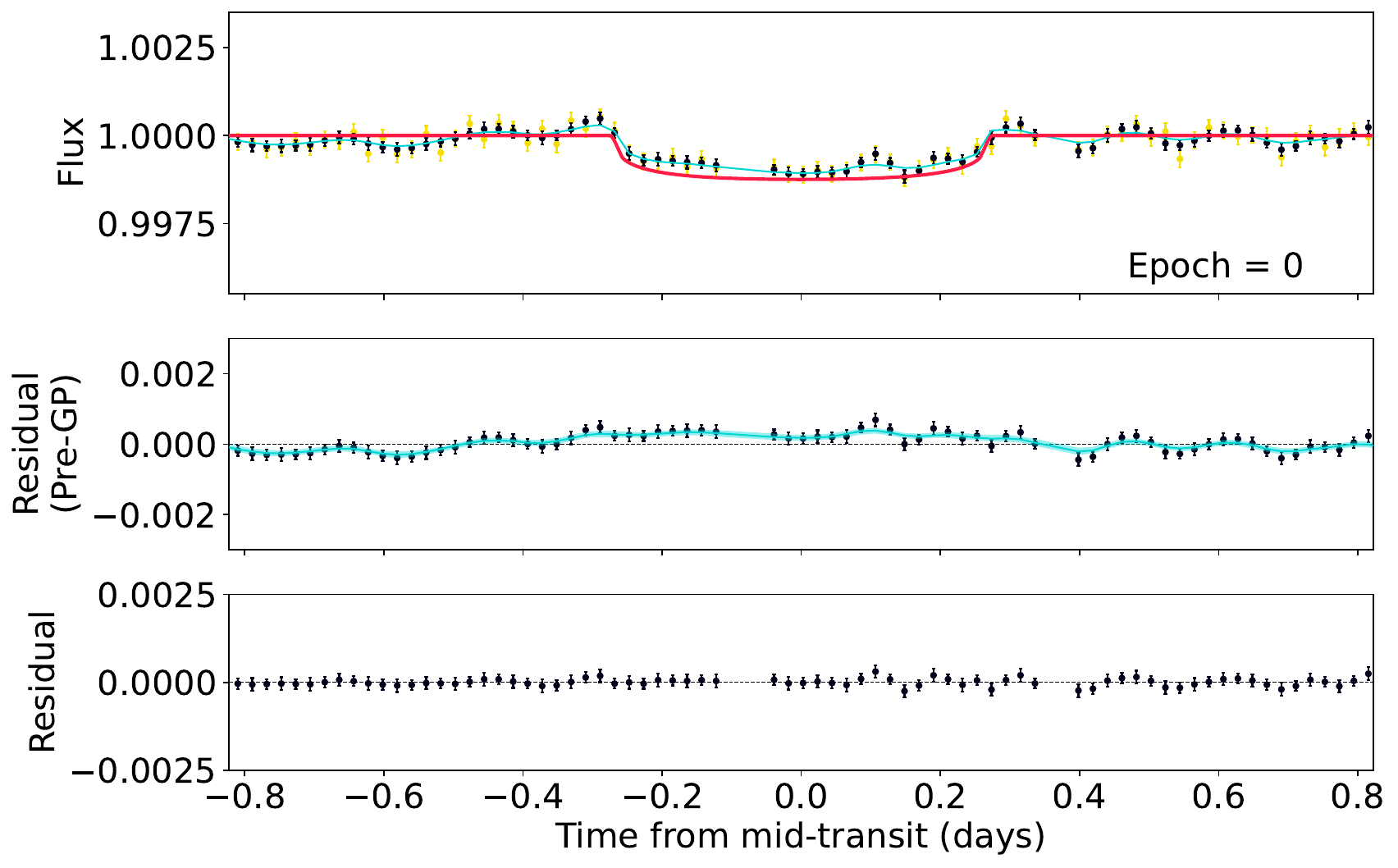}\vspace{0.3cm} &
    \includegraphics[width=0.48\linewidth]{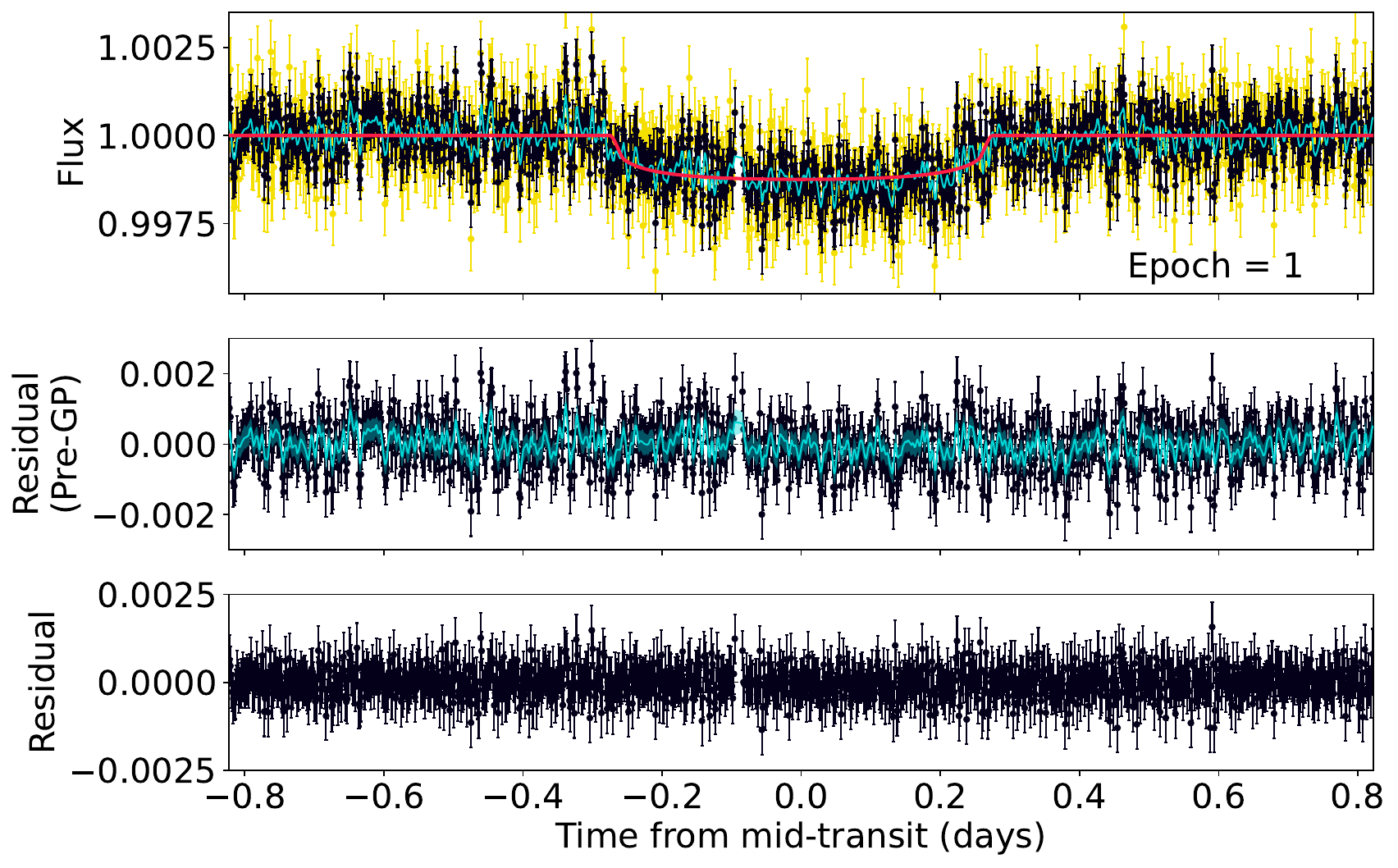}\vspace{0.3cm} \\
    \includegraphics[width=0.48\linewidth]{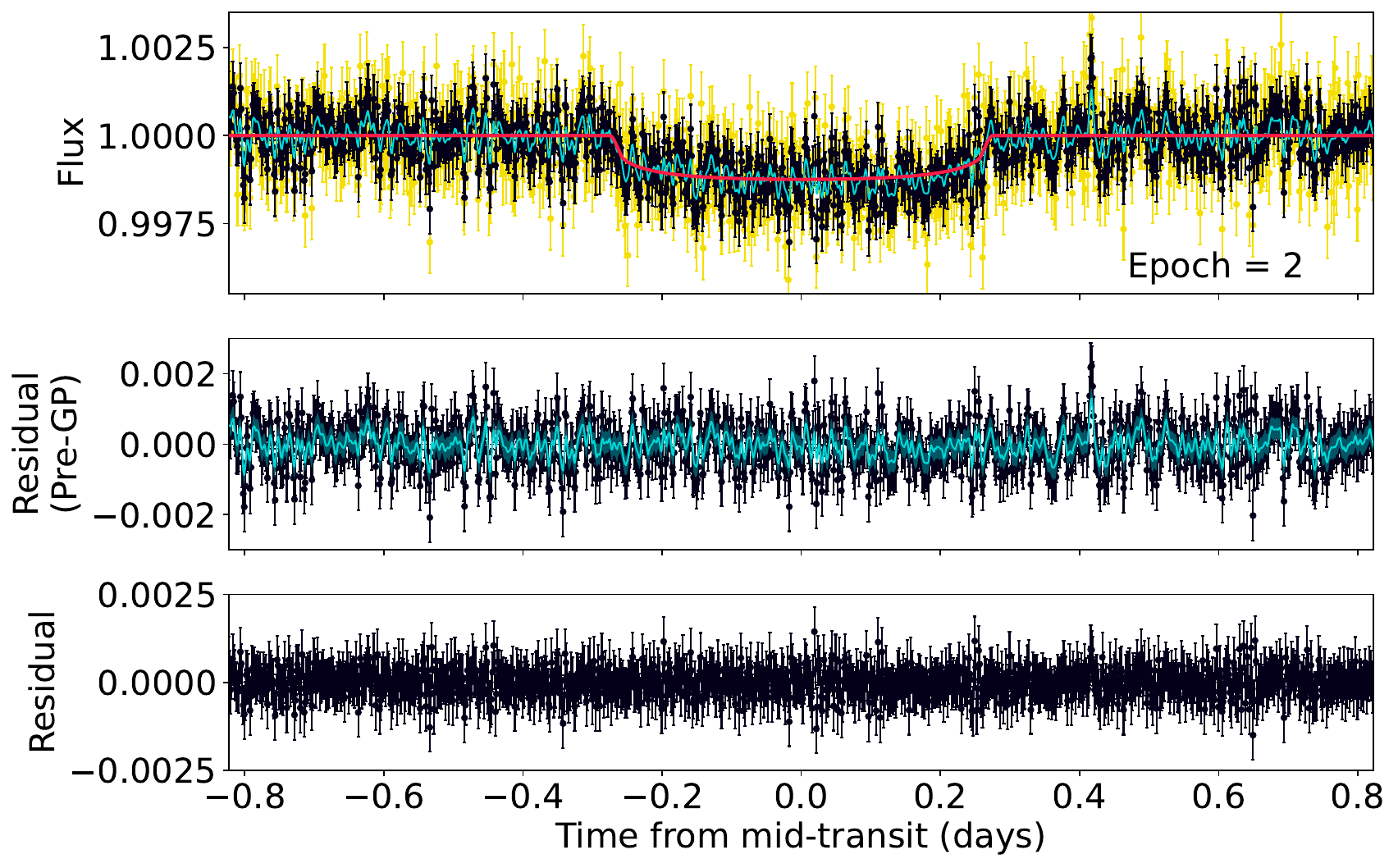}\vspace{0.3cm} &
    \includegraphics[width=0.48\linewidth]{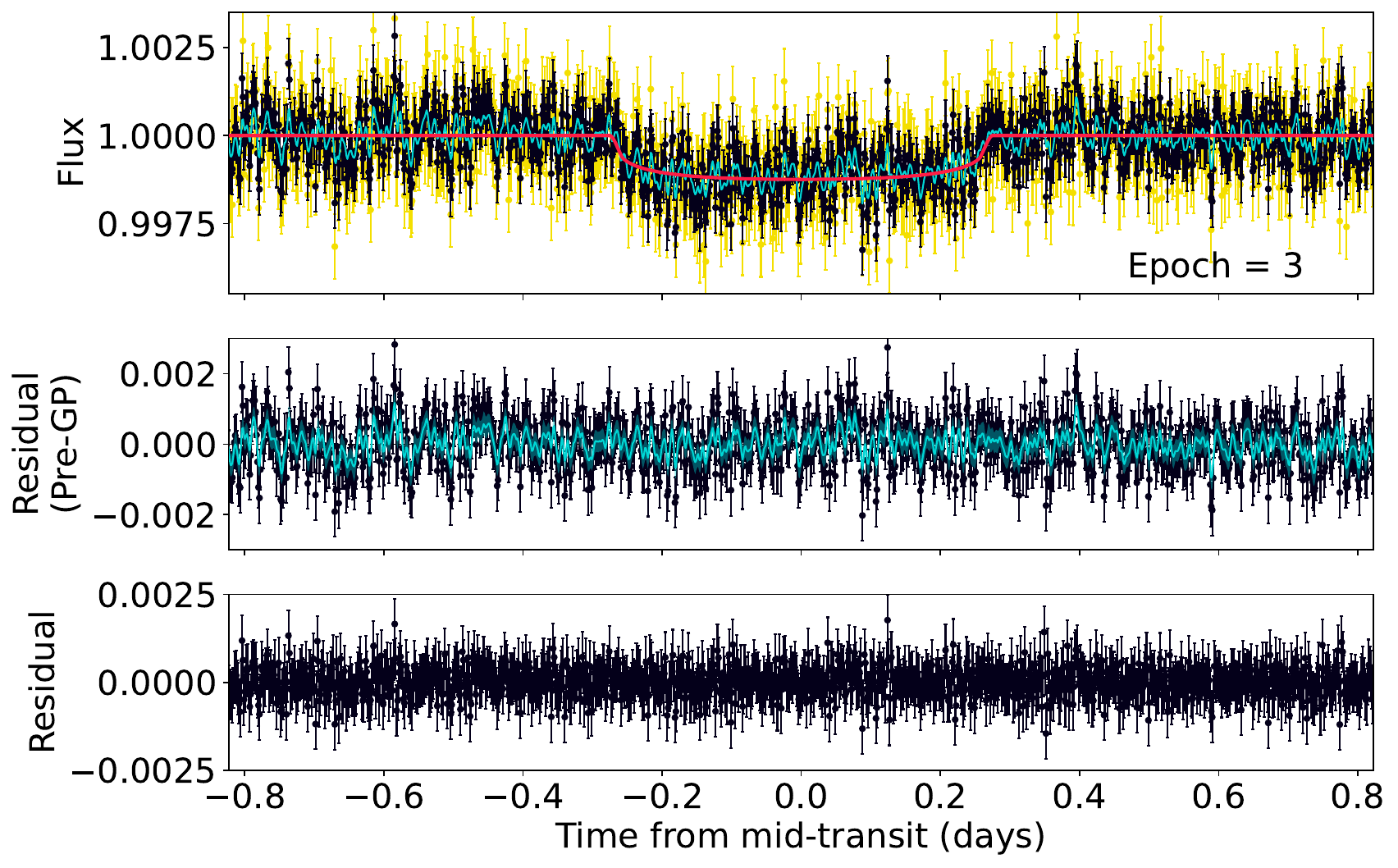}\vspace{0.3cm}\\
    \includegraphics[width=0.48\linewidth]{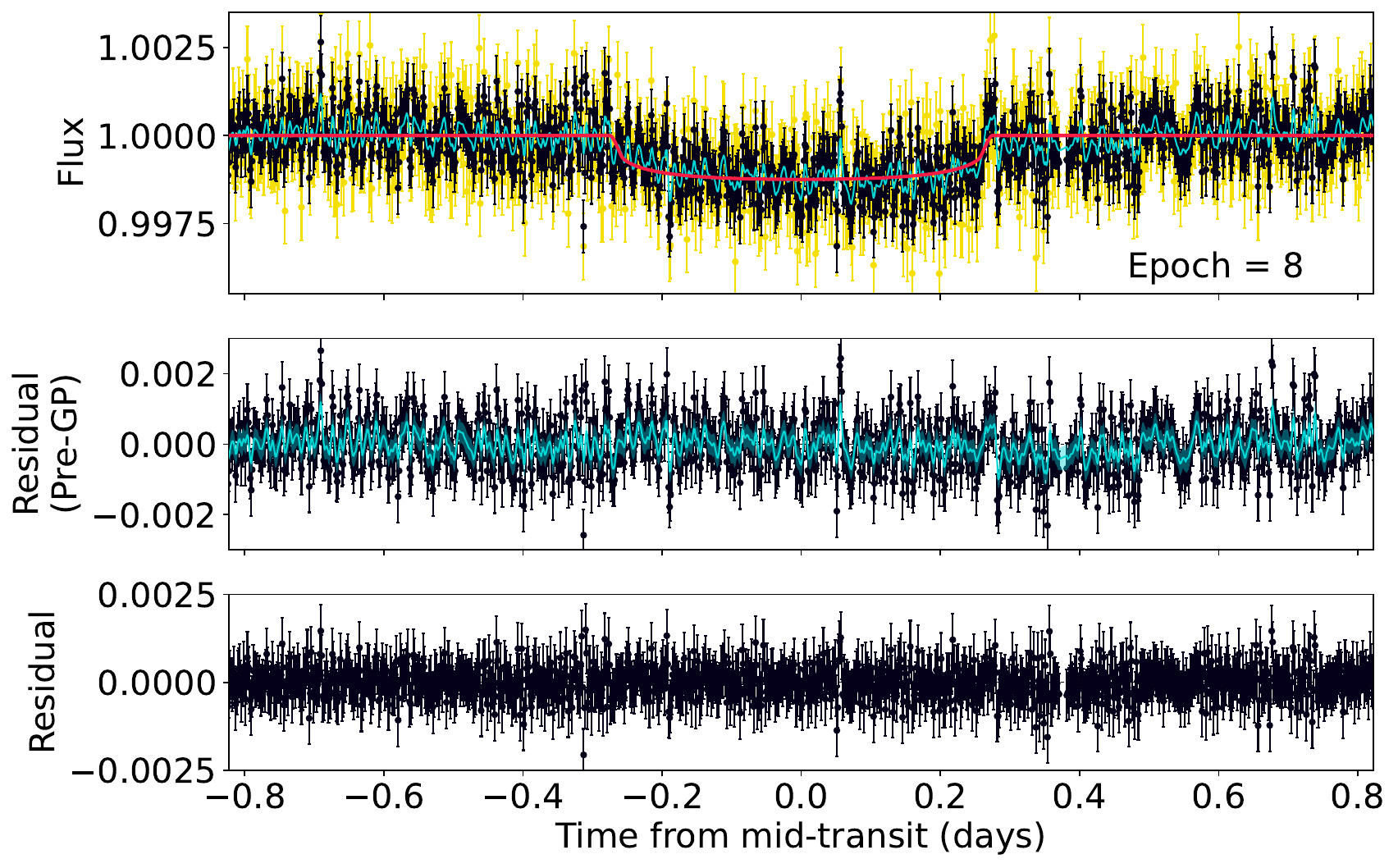}\vspace{0.3cm} &
    \includegraphics[width=0.48\linewidth]{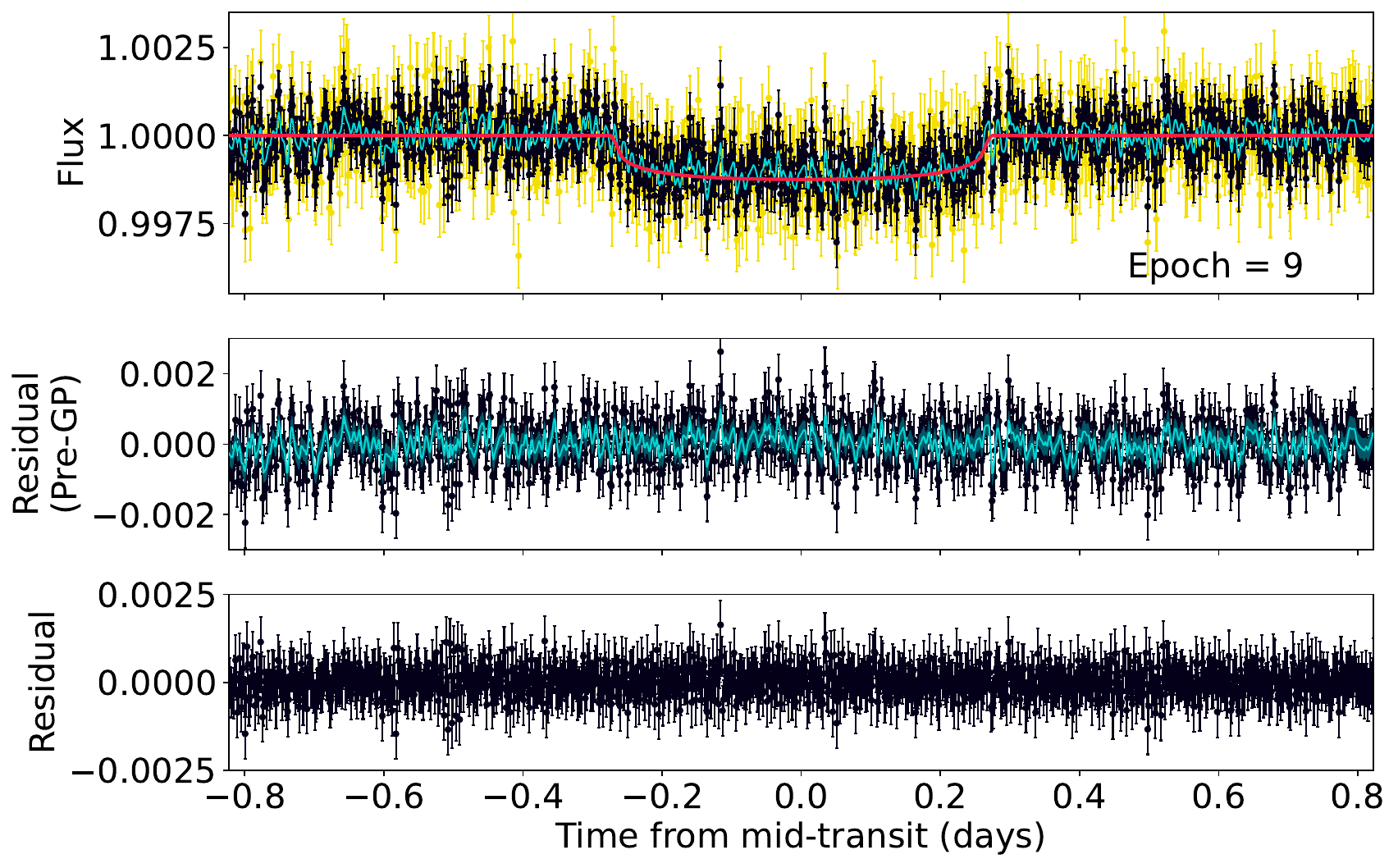}\vspace{0.3cm}\\
    \includegraphics[width=0.48\linewidth]{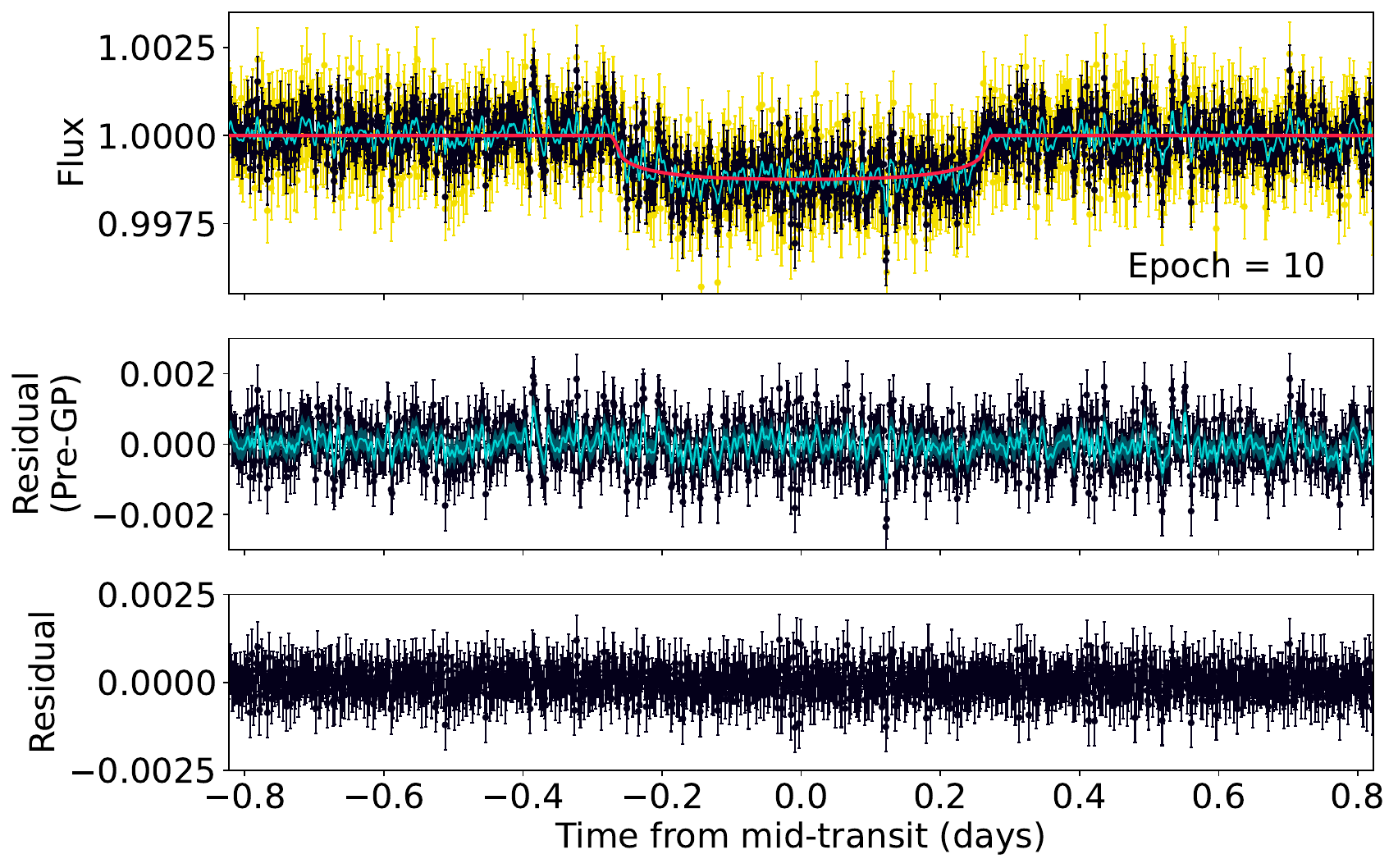}\vspace{0.2cm} &
    \includegraphics[width=0.48\linewidth]{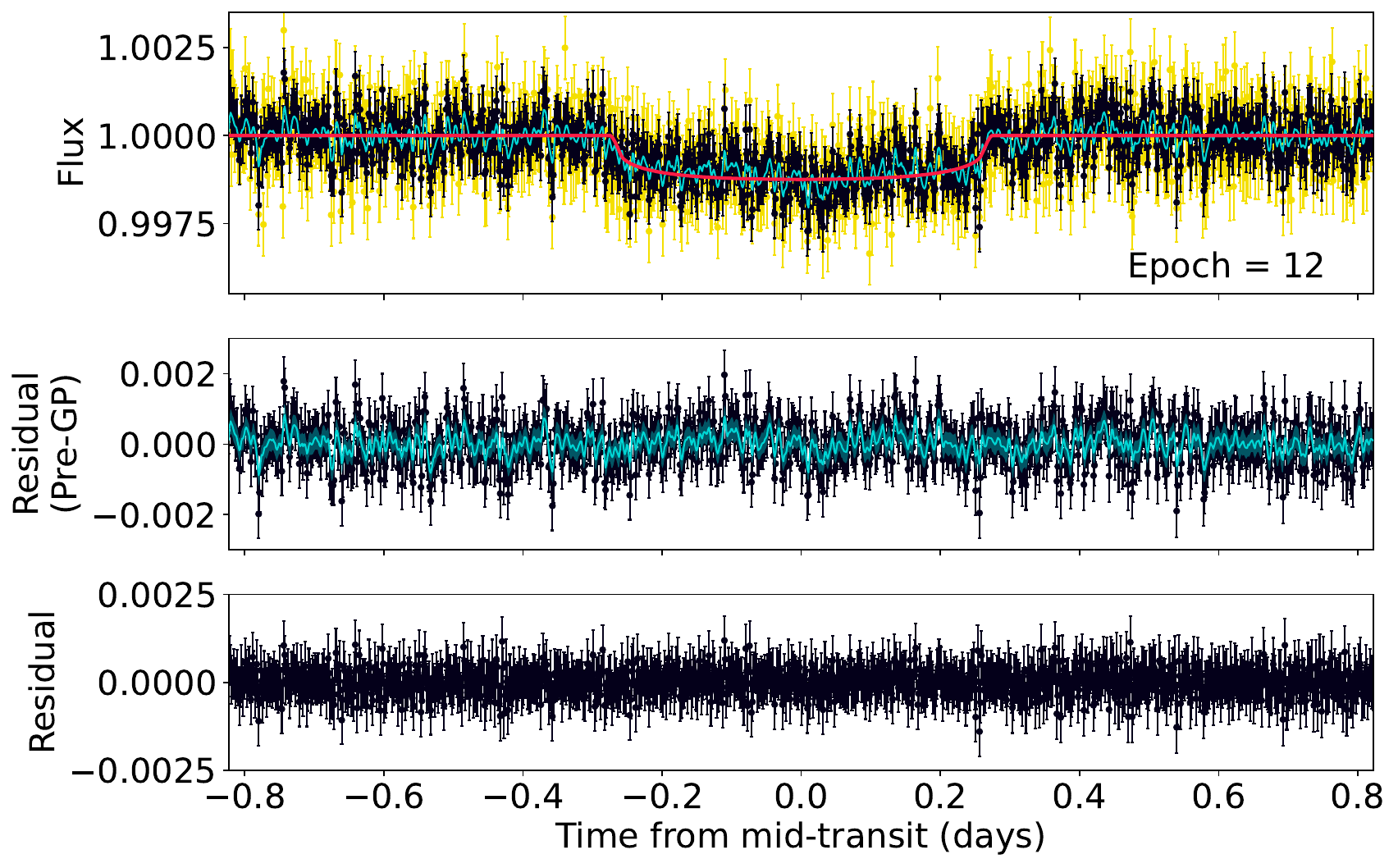}\vspace{0.2cm}
    \end{tabular}
	\caption{Same as Figure \ref{fig:fig1}, but for TOI-813 b (part 1 of 2).}
	\label{fig:fig5}
    \vspace{0.3cm}
\end{figure*}

\addtocounter{figure}{-1}

\begin{figure*}[!t]
	\centering
    \begin{tabular}{cc}
    \includegraphics[width=0.48\linewidth]{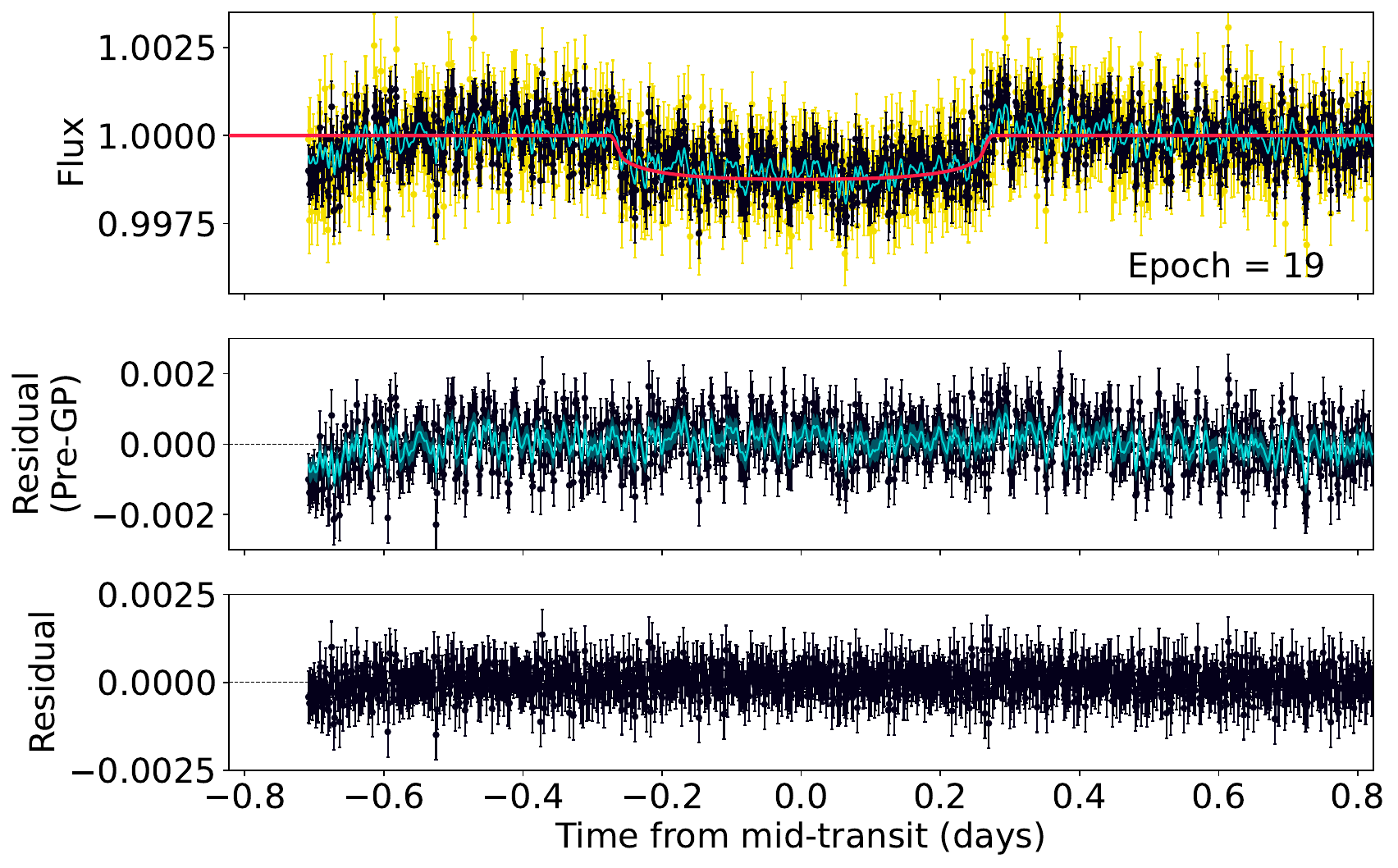}\vspace{0.3cm} &
    \includegraphics[width=0.48\linewidth]{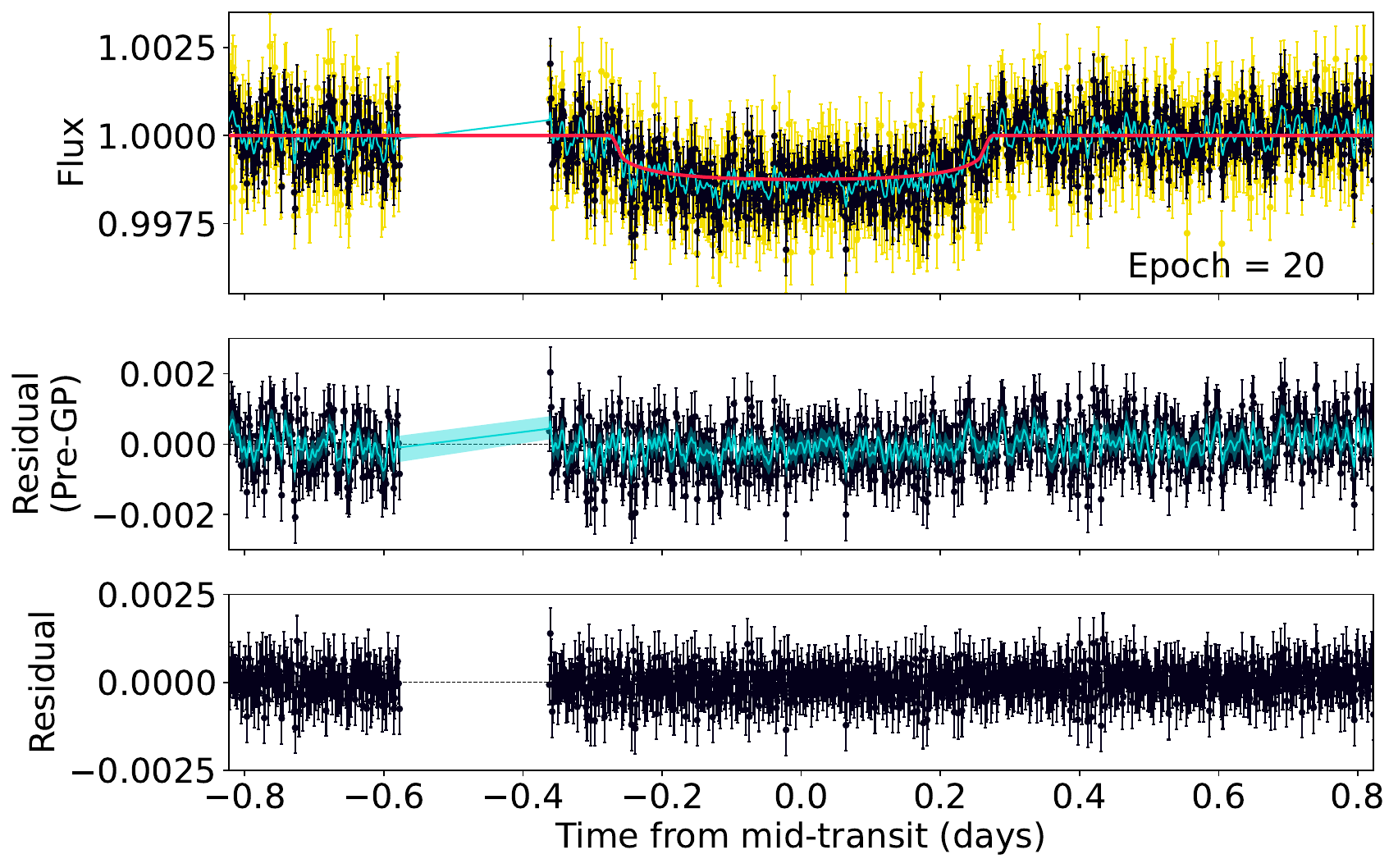}\vspace{0.3cm} \\
    \multicolumn{2}{c}{\includegraphics[width=0.48\linewidth]{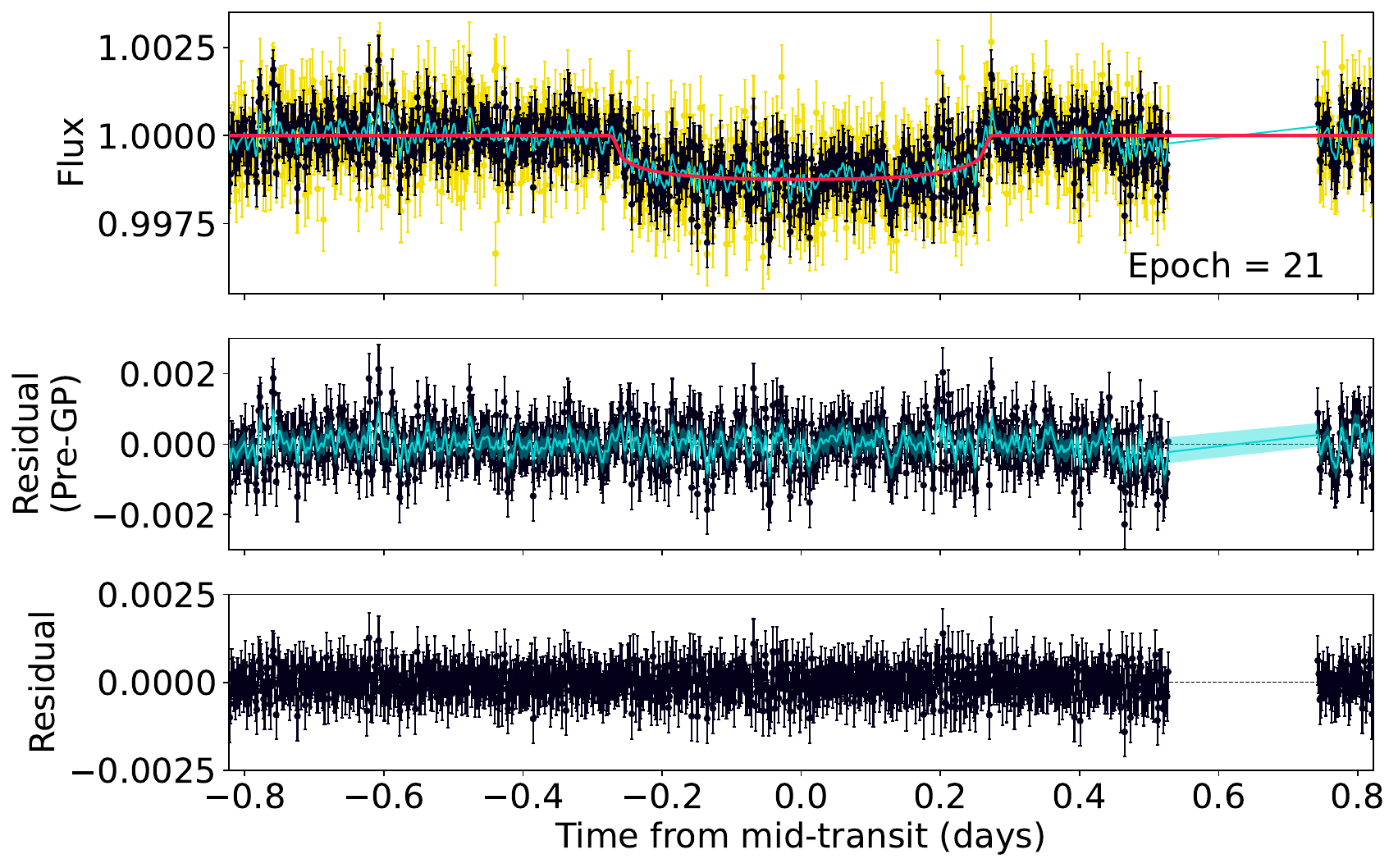}\vspace{0.2cm}}
    \end{tabular}
	\caption{(continued, part 2 of 2).}
\end{figure*}

In contrast to these relative advantages of the long-period transiting exoplanets, their known population is significantly small, as the transit probability decreases with the increase in the orbital distances. Ideally, it requires the detection of subsequent transit events to verify the nature and the orbital properties of any transiting exoplanet candidate. However, because of the intrinsic nature of the long-period exoplanets, the subsequent detection of their transit events would require a much longer baseline of observations. In the past, only the Kepler \citep{2010Sci...327..977B} mission was able to detect several long-period exoplanet candidates through subsequent transit detections. This was possible because of the very long continuous baseline of Kepler. However, since the field of view of Kepler is limited, and because of the relative scarcity of long-period transiting exoplanets, the majority of these candidates detected by Kepler were found around comparatively fainter stars. This, in fact, does not make them the most suitable candidates for the follow-up studies discussed previously using the existing facilities, which is also evident from the relative scarcity of such studies at present.

In recent years, the Transiting Exoplanet Survey Satellite \cite[TESS,][]{2015JATIS...1a4003R} has been able to detect several long-period exoplanet candidates around very bright sources, which makes it possible to verify them through radial velocity and transit follow-up observations using ground-based instruments. This has resulted in several such confirmed long-period exoplanets, making them among the most interesting targets for future follow-up studies. However, since the transit frequency of these exoplanets is very low, many of these confirmed planets have only very few reported transits in the literature. The known ephemeris of several of these exoplanets have large uncertainties, which make it difficult to plan precise follow-up observations. Also, data from fewer transits could result in an inaccurate estimation of the physical parameters originating from small data biases. As these physical properties are directly adopted in several follow-up studies \citep[e.g. for WASP-79 b:][]{2020AJ....160..109S, 2022MNRAS.514.5192L, 2022AJ....163....7F}, a reliable estimation of these properties using a larger volume of transit photometric data is necessary.

In this work, I aim to use the newly available data for a few of these interesting long-period exoplanets around bright sources to improve their ephemeris and update their physical properties. The publicly available follow-up observations from TESS and CHEOPS \citep{2021ExA....51..109B} were used, and a well-tested critical noise treatment algorithm was used to analyze them. In section \ref{sec:sec2}, I have discussed the targets and the data used from the study. In section \ref{sec:sec3}, I have described the data analyses and modeling algorithms. And finally, in section \ref{sec:sec4}, I have discussed the results obtained from this work.

\section{Targets and Observations}\label{sec:sec2}

For this work, the targets were selected from the pool of known long-period transiting exoplanets, having newer publicly available observations from TESS and/or CHEOPS, with a priority to those with relatively large uncertainties in their known ephemeris. The search resulted in a few very interesting targets, from which the five most interesting targets from the perspective of future follow-up studies, i.e., with very bright sources (V $<$ 11), were selected for this work. They are: HD95338 b, TOI-2134 c, K2-290 c, TOI-1898 b, and TOI-813 b. Among these, HD95338 b is a Neptune-sized planet, TOI-2134 c and TOI-813 b are sub-Saturns, and K2-290 c and TOI-1898 b are Jupiter-sized.

HD95338 b was first reported by \cite{2020MNRAS.496.4330D}, consisting of only a single transit detection by TESS in sector 10. HD95338 has been observed by TESS over four more sectors, from which I have identified two additional transit events in sectors 36 and 63 during epochs 13 and 26, respectively. Additionally, CHEOPS has observed three transits of HD95338 b corresponding to epochs 13, 20, and 27 (PI: S Charnoz, A Simon). TOI-2134 c was first reported by \cite{2024MNRAS.527.5385R}, which also consisted of only a single transit detection by TESS in sector 52. The subsequent observations by TESS have resulted in the detection of another transit in sector 80, corresponding to epoch 8. K2-290 c was previously reported by \cite{2019MNRAS.484.3522H} with two observed transits by K2. There has been no TESS follow-up of K2-290 yet, however, CHEOPS has observed two transits of K2-290 c, corresponding to epochs 35 and 42 (PI: S Charnoz, A Simon). For TOI-1898b, three transits were previously reported by \cite{2024arXiv240207893C} in the sectors 21, 45, and 46 of TESS, corresponding to epochs 0, 14, and 15, respectively. Subsequently, three additional transits of TOI-1898b have been observed by CHEOPS, corresponding to epochs 23, 24, and 25 (PI: A Simon). Four transits of TOI-813 b, in sectors 2, 5, 8, and 11 of TESS, were previously reported by \cite{2020MNRAS.494..750E}, corresponding to epochs 0, 1, 2, and 3, respectively. From the subsequent observations by TESS, I have identified seven additional transits in sectors 27, 30, 33, 39, 61, 64, and 67, corresponding to epochs 8, 9, 10, 12, 19, 20, and 21.

All the publicly available data from TESS and K2 were obtained using the Barbara A. Mikulski Archive for Space Telescopes (MAST\footnote{https://mast.stsci.edu/}), and those from CHEOPS were obtained using the official CHEOPS Archive Browser\footnote{https://cheops-archive.astro.unige.ch/archive$\_$browser/}.

\section{Data analysis and Modeling}\label{sec:sec3}

For most of the TESS observations used in this work, the 120s cadence SPOC/PDCSAP lightcurves were available from MAST, which were used in further analyses. Only for sector 36 of HD95338 b and sector 2 of TOI-813 b, where the SPOC lightcurves were unavailable, the lower cadence QLP lightcurves were used. For the K2 observations, the 1800s cadence PDCSAP lightcurves were used. Lower cadence data can introduce biases in the parameter estimation; however, since they were modeled simultaneously with multiple higher cadence datasets for each target, any potential biases were effectively nullified. The CHEOPS data products downloaded from the official archive consist of lightcurves reduced using the official Data Reduction Pipeline (DRP), with the apertures ranging between 15-40. The lightcurves corresponding to the aperture 24 were used in further analyses, as they were found to be the best in terms of lightcurve dispersion for the majority of the cases.

\begin{table}
    \centering
    \caption{Estimated physical properties of HD95338 b.}
    \label{tab:tab1}
    $\begin{array}{lc}
    \hline
    \hline
    \text{Parameter} & \text{Value}\\
    \hline
    \text{Transit parameters} &\\
    T_0\;[BJD_{TDB}] & 2458585.28072_{-0.00054}^{+0.00055}\\
    P \;[days] & 55.082695\pm0.000027\\
    b & 0.25_{-0.15}^{+0.11}\\
    R_\star/a & 0.01447_{-0.00037}^{+0.00051}\\
    R_p/R_{\star, TESS} & 0.04261_{-0.00051}^{+0.00057}\\
    R_p/R_{\star, CHEOPS} & 0.04369_{-0.00041}^{+0.00049}\\
    \hline
    \text{Limb-darkening coefficients} & \\
    C_{1, TESS} & 0.536_{-0.097}^{+0.091}\\
    C_{2, TESS} & 0.22\pm0.14\\
    C_{1, CHEOPS} & 0.413_{-0.083}^{+0.095}\\
    C_{2, CHEOPS} & 0.42_{-0.16}^{+0.14}\\
    \hline
    \text{Derived parameters} & \\
    T_{14}\;[hr] & 5.613_{-0.024}^{+0.03}\\
    a/R_\star & 69.1_{-2.3}^{+1.8}\\
    i \;[deg] & 89.77_{-0.11}^{+0.14}\\
    T_{eq}\;[K] & 443_{-5.7}^{+7.8}\\
    a\;[AU] & 0.279\pm0.015\\
    M_p\;[M_J] & 0.135_{-0.0067}^{+0.0068}\\
    M_p\;[M_\oplus] & 42.9_{-2.1}^{+2.2}\\
    R_{p, TESS}\;[R_J] & 0.361\pm0.017\\
    R_{p, TESS}\;[R_\oplus] & 4.04\pm0.19\\
    R_{p, CHEOPS}\;[R_J] & 0.37\pm0.017\\
    R_{p, CHEOPS}\;[R_\oplus] & 4.15\pm0.19\\
    \hline
    \end{array}$
\end{table}

\begin{table}
    \centering
    \caption{Estimated physical properties of TOI-2134 c.}
    \label{tab:tab2}
    $\begin{array}{lc}
    \hline
    \hline
    \text{Parameter} & \text{Value}\\
    \hline
    \text{Transit parameters} & \\
    T_0\;[BJD_{TDB}] & 2459718.96933\pm0.00026\\
    P \;[days] & 95.852876_{-0.000051}^{+0.000049}\\
    b & 0.417_{-0.048}^{+0.042}\\
    R_\star/a & 0.01427_{-0.0003}^{+0.00032}\\
    R_p/R_\star & 0.0935_{-0.00069}^{+0.00073}\\
    \hline
    \text{Limb-darkening coefficients} & \\
    C_1 & 0.537_{-0.076}^{+0.047}\\
    C_2 & 0.101_{-0.068}^{+0.137}\\
    \hline
    \text{Derived parameters} & \\
    T_{14}\;[hr] & 5.263_{-0.027}^{+0.029}\\
    a/R_\star & 70.1\pm1.5\\
    i \;[deg] & 89.08\pm0.12\\
    T_{eq}\;[K] & 386.8_{-5.9}^{+6.1}\\
    a\;[AU] & 0.231_{-0.0074}^{+0.0075}\\
    M_p\;[M_J] & 0.18\pm0.03\\
    M_p\;[M_\oplus] & 57.2\pm9.4\\
    R_p\;[R_J] & 0.645\pm0.016\\
    R_p\;[R_\oplus] & 7.23\pm0.18\\
    \hline
    \end{array}$
\end{table}

\begin{table}
    \centering
    \caption{Estimated physical properties of K2-290 c.}
    \label{tab:tab3}
    $\begin{array}{lc}
    \hline
    \hline
    \text{Parameter} & \text{Value}\\
    \hline
    \text{Transit parameters} & \\
    T_0\;[BJD_{TDB}] & 2458019.17235\pm0.00034\\
    P \;[days] & 48.367761\pm0.000042\\
    b & 0.557_{-0.035}^{+0.029}\\
    R_\star/a & 0.02461_{-0.00058}^{+0.00055}\\
    R_p/R_{\star, K2} & 0.06875_{-0.00067}^{+0.00055}\\
    R_p/R_{\star, CHEOPS} & 0.06778_{-0.0009}^{+0.00102}\\
    \hline
    \text{Limb-darkening coefficients} & \\
    C_{1, K2} & 0.339_{-0.133}^{+0.098}\\
    C_{2, K2} & 0.29_{-0.15}^{+0.2}\\
    C_{1, CHEOPS} & 0.35_{-0.18}^{+0.13}\\
    C_{2, CHEOPS} & 0.24_{-0.16}^{+0.21}\\
    \hline
    \text{Derived parameters} & \\
    T_{14}\;[hr] & 8.281_{-0.034}^{+0.038}\\
    a/R_\star & 40.64_{-0.88}^{+0.97}\\
    i \;[deg] & 89.214_{-0.059}^{+0.068}\\
    T_{eq}\;[K] & 699\pm16\\
    a\;[AU] & 0.286\pm0.015\\
    M_p\;[M_J] & 0.774_{-0.044}^{+0.045}\\
    M_p\;[M_\oplus] & 246\pm14\\
    R_{p, K2}\;[R_J] & 1.01\pm0.049\\
    R_{p, K2}\;[R_\oplus] & 11.32\pm0.55\\
    R_{p, CHEOPS}\;[R_J] & 0.997_{-0.049}^{+0.05}\\
    R_{p, CHEOPS}\;[R_\oplus] & 11.17_{-0.55}^{+0.56}\\
    \hline
    \end{array}$
\end{table}

The inhomogeneous and asymmetric shape of the CHEOPS point spread function adds a roll-angle-dependent variation from the background flux to the lightcurves \citep{2021ExA....51..109B, 2023A&A...675A..81H}. It was modeled using a glint function and subtracted from the lightcurves using PyCHEOPS \citep{2022MNRAS.514...77M}. For easier handling of the time-correlated noise in the lightcurves, the long TESS lightcurves were sliced into smaller sections consisting of individual transits, i.e., the transit lightcurves. Both TESS and CHEOPS transit lightcurves were baseline corrected by modeling the out-of-transit section using a first/second order polynomial, and then subtracting the best-fit model from the entire lightcurves, determined by least Bayesian Information Criterion \citep[BIC,][]{1978AnSta...6..461S}.

The lightcurves were treated using the wavelet denoising \citep{Donoho1994IdealDI, 806084, WaveletDenoise2012, 2021AJ....162...18S, 2023ApJS..268....2S, 2024ApJS..274...13S} technique in order to reduce the fluctuations due to background variations, uncorrelated in time. Wavelet denoising does not affect the higher frequency components in the signal and thus is considered a better alternative to other smoothing techniques like binning or Gaussian moving average. Due to the very low cadence of the K2 lightcurves, wavelet denoising was not performed on them. For all the rest of the lightcurves, only first-order wavelet denoising was performed to avoid oversmoothing. The PyWavelets \citep{Lee2019} package was used for the discrete wavelet transforms (DWT), along with the Symlet family of wavelets \citep{daubechies1988orthonormal}, and the Universal Thresholding Law \citep{Donoho1994IdealDI}.

In order to deal with the time-correlated noise in the lightcurves, the Gaussian process (GP) regression \citep{2006gpml.book.....R, 2015ApJ...810L..23J, 2019MNRAS.489.5764P, 2021AJ....162..221S, 2023ApJS..268....2S, 2024ApJS..274...13S} technique was used. GP regression was applied simultaneously with the transit lightcurves modeling, thereby modeling the time-correlated noise components in the lightcurves in parallel to modeling the transit signals. The Matern class covariance function with $\nu$ = 3/2 was used, with two free parameters, i.e., the signal standard deviation ($\alpha$) and characteristic time scale ($\tau$). Separate GP coefficients were modeled for each telescope and lightcurve cadence. The transit parameters were modeled using the analytical transit formalism by \citet{2002ApJ...580L.171M}, with the application of quadratic limb darkening law. The Markov chain Monte Carlo (MCMC) sampling technique was used for the modeling, with the implementation of the Hastings-Metropolis algorithm \citep{1970Bimka..57...97H}. Uniform priors were used for each parameter, with wide enough ranges to allow for bias-free convergence.

The wavelength-dependent parameters, such as $R_p/R\star$ and the limb-darkening coefficients, were modeled independently for each telescope. The orbital eccentricity and the argument of the periapses, which are estimated from the radial velocity (RV) measurements, were fixed to the median values from the literature. The RV semi-amplitude and the properties of the host stars were also adopted from the literature for the calculation of other derivable parameters. Figures \ref{fig:fig1}-\ref{fig:fig5} show all the observed and model transit lightcurves, highlighting the impact of the wavelet denoising and the GP regression techniques. Tables \ref{tab:tab1}-\ref{tab:tab5} present all the estimated parameters for the targets from this work. Table \ref{tab:gp} shows the estimated GP regression coefficients.

\begin{table}
    \centering
    \caption{Estimated physical properties of TOI-1898 b.}
    \label{tab:tab4}
    $\begin{array}{lc}
    \hline
    \hline
    \text{Parameter} & \text{Value}\\
    \hline
    \text{Transit parameters} & \\
    T_0\;[BJD_{TDB}] & 2458894.25409_{-0.00052}^{+0.00051}\\
    P \;[days] & 45.522129_{-0.000023}^{+0.000024}\\
    b & 0.773_{-0.016}^{+0.011}\\
    R_\star/a & 0.02921_{-0.00074}^{+0.00061}\\
    R_p/R_{\star, TESS} & 0.05497_{-0.00052}^{+0.0005}\\
    R_p/R_{\star, CHEOPS} & 0.05519_{-0.00049}^{+0.00042}\\
    \hline
    \text{Limb-darkening coefficients} & \\
    C_{1, TESS} & 0.23_{-0.18}^{+0.17}\\
    C_{2, TESS} & 0.28_{-0.17}^{+0.19}\\
    C_{1, CHEOPS} & 0.21_{-0.14}^{+0.16}\\
    C_{2, CHEOPS} & 0.39_{-0.17}^{+0.16}\\
    \hline
    \text{Derived parameters} & \\
    T_{14}\;[hr] & 4.375_{-0.023}^{+0.022}\\
    a/R_\star & 34.23_{-0.7}^{+0.89}\\
    i \;[deg] & 87.531_{-0.089}^{+0.113}\\
    T_{eq}\;[K] & 754_{-14}^{+13}\\
    a\;[AU] & 0.257_{-0.0072}^{+0.008}\\
    M_p\;[M_J] & 0.467\pm0.023\\
    M_p\;[M_\oplus] & 148.3\pm7.4\\
    R_{p, TESS}\;[R_J] & 0.863\pm0.017\\
    R_{p, TESS}\;[R_\oplus] & 9.67\pm0.19\\
    R_{p, CHEOPS}\;[R_J] & 0.866\pm0.017\\
    R_{p, CHEOPS}\;[R_\oplus] & 9.71\pm0.19\\
    \hline
    \end{array}$
\end{table}

\begin{table}
    \centering
    \caption{Estimated physical properties of TOI-813 b}
    \label{tab:tab5}
    $\begin{array}{lc}
    \hline
    \hline
    \text{Parameter} & \text{Value}\\
    \hline
    \text{Transit parameters} & \\
    T_0\;[BJD_{TDB}] & 2458370.7706\pm0.003\\
    P \;[days] & 83.89608\pm0.00024\\
    b & 0.24_{-0.16}^{+0.18}\\
    R_\star/a & 0.02048_{-0.00059}^{+0.0015}\\
    R_p/R_\star & 0.03225_{-0.00062}^{+0.00064}\\
    \hline
    \text{Limb-darkening coefficients} & \\
    C_1 & 0.28_{-0.15}^{+0.14}\\
    C_2 & 0.48\pm0.3\\
    \hline
    \text{Derived parameters} & \\
    T_{14}\;[hr] & 13.16_{-0.15}^{+0.2}\\
    a/R_\star & 48.8_{-3.3}^{+1.5}\\
    i \;[deg] & 89.72_{-0.25}^{+0.18}\\
    T_{eq}\;[K] & 601_{-19}^{+23}\\
    a\;[AU] & 0.436_{-0.033}^{+0.029}\\
    R_p\;[R_J] & 0.609_{-0.033}^{+0.034}\\
    R_p\;[R_\oplus] & 6.82_{-0.37}^{+0.38}\\
    \hline
    \end{array}$
\end{table}

\begin{table}
    \centering
    \caption{Estimated values of GP regression coefficients.}
    \label{tab:gp}
    $\begin{array}{lccc}
        \hline
        \hline
         \text{Target} & \text{Case} & \text{$\alpha$} & \text{$\tau$ [days]} \\
         \hline
        \text{HD95338 b} & \text{TESS} & 1.63_{-0.28}^{+0.29} \times 10^{-4} & 3.49_{-0.67}^{+1.10} \times 10^{-3} \\
        & \text{TESS(LC)} & 1.03_{-0.28}^{+0.45} \times 10^{-4} & 1.72_{-0.60}^{+1.03} \times 10^{-2} \\
        & \text{CHEOPS} & 1.44_{-0.08}^{+0.08} \times 10^{-4} & 6.3_{-0.5}^{+0.5} \times 10^{-4} \\
        \text{TOI-2134 b} & \text{TESS} & 2.12_{-0.12}^{+0.12} \times 10^{-4} & 3.04_{-0.25}^{+0.29} \times 10^{-3} \\
        \text{TOI-290 c} & \text{K2} & 6.6_{-0.7}^{+0.9} \times 10^{-5} & 1.44_{-0.64}^{+0.47} \times 10^{-2} \\
        & \text{CHEOPS} & 1.22_{-0.07}^{+0.08} \times 10^{-3} & 1.33_{-0.09}^{+0.08} \times 10^{-3} \\
        \text{TOI-1898 b} & \text{TESS} & 1.80_{-0.11}^{+0.11} \times 10^{-4} & 2.65_{-0.20}^{+0.24} \times 10^{-3} \\
        & \text{CHEOPS} & 9.9_{-0.4}^{+0.5} \times 10^{-5} & 7.7_{-0.7}^{+0.8} \times 10^{-4} \\
        \text{TOI-813 b} & \text{TESS(LC)} & 1.86_{-0.36}^{+0.36} \times 10^{-4} & 6.59_{-1.62}^{+2.69} \times 10^{-2} \\
        & \text{TESS} & 4.7_{-0.5}^{+0.5} \times 10^{-4} & 2.58_{-0.35}^{+0.57} \times 10^{-3} \\
        \hline
    \end{array}$
\end{table}

\begin{table*}
    \centering
    \caption{Comparison of estimated parameters with the previous studies involving observations from space-based instruments.}
    \label{tab:tab6}
    $\begin{array}{lcccccc}
        \hline
        \hline
         \text{Target} & \text{Study} & \text{$T_0[BJD_{TDB}]$} & \text{$P$ [days]} & \text{$b$} & \text{$a/R_\star$} & \text{$R_{p}/R_\star$}\\
         \hline
        \text{HD95338 b} & \text{This Work} & 2458585.28072_{-0.00054}^{+0.00055} & 55.082695\pm0.000027 & 0.25_{-0.15}^{+0.11} & 69.1_{-2.3}^{+1.8} & 0.04261_{-0.00051}^{+0.00057\,\dag}\\
        & \text{\cite{2020MNRAS.496.4330D}} & 2458585.2795\pm0.0006 & 55.087_{-0.02}^{+0.02} & 0.43_{-0.113}^{+0.07} & 64.676_{-3.84}^{+3.81} & 0.0401\pm0.0027\\
        \text{TOI-2134 b} & \text{This Work} & 2459718.96933\pm0.00026 & 95.852876_{-0.000051}^{+0.000049} & 0.417_{0.048}^{+0.042} & 70.1\pm1.5 & 0.0935_{-0.00069}^{+0.00073}\\
        & \text{\cite{2024MNRAS.527.5385R}} & 2459718.96939\pm0.00020 & 95.5_{-0.25}^{+0.36} & 0.464\pm0.042 & 112\pm2 & 0.09404\pm0.00078\\
        \text{K2-290 c} & \text{This Work} & 2458019.17235\pm0.00034 & 48.367761\pm0.000042 & 0.557_{-0.035}^{+0.029} & 40.64_{-0.88}^{+0.97} & 0.06875_{-0.00067}^{+0.00055\,\ddag}\\
        & \text{\cite{2019MNRAS.484.3522H}} & 2458019.17333\pm0.00029 & 48.36685_{-0.0004}^{+0.00041} & 0.474\pm0.062 & 43.5\pm1.2 & 0.06848_{-0.00047}^{+0.00042}\\
        \text{TOI-1898 b} & \text{This Work} & 2458894.25409_{-0.00052}^{+0.00051} & 45.522129_{-0.000023}^{+0.000024} & 0.773_{-0.016}^{+0.011} & 34.23_{-0.7}^{+0.89} & 0.05497_{-0.00052}^{+0.0005\,\dag}\\
        & \text{\cite{2024ApJS..272...32P}} & 2458894.2511\pm0.0008 & 45.52234\pm0.000061 & 0.7_{-0.112}^{+0.054} & 35.85\pm0.95 & 0.0534_{-0.0014}^{+0.0009}\\
        \text{TOI-813 b} & \text{This Work} & 2458370.7706\pm0.003 & 83.89608\pm0.00024 & 0.24_{-0.16}^{+0.18} & 48.8_{-3.3}^{+1.5} & 0.03225_{-0.00062}^{+0.00064}\\
        & \text{\cite{2020MNRAS.494..750E}} & 2458370.7836_{-0.0062}^{+0.0072} & 83.8911_{-0.0031}^{+0.0027} & 0.3_{-0.19}^{+0.18} & 47.2_{-3.8}^{+2.1} & 0.03165_{-0.00061}^{+0.00072}\\
        \hline
        \multicolumn{6}{l}{\text{$^\dag$ Corresponding to TESS observations}} \\
        \multicolumn{6}{l}{\text{$^\ddag$ Corresponding to K2 observations}}
    \end{array}$
\end{table*}

\begin{figure}
	\centering
	\includegraphics[width=\linewidth]{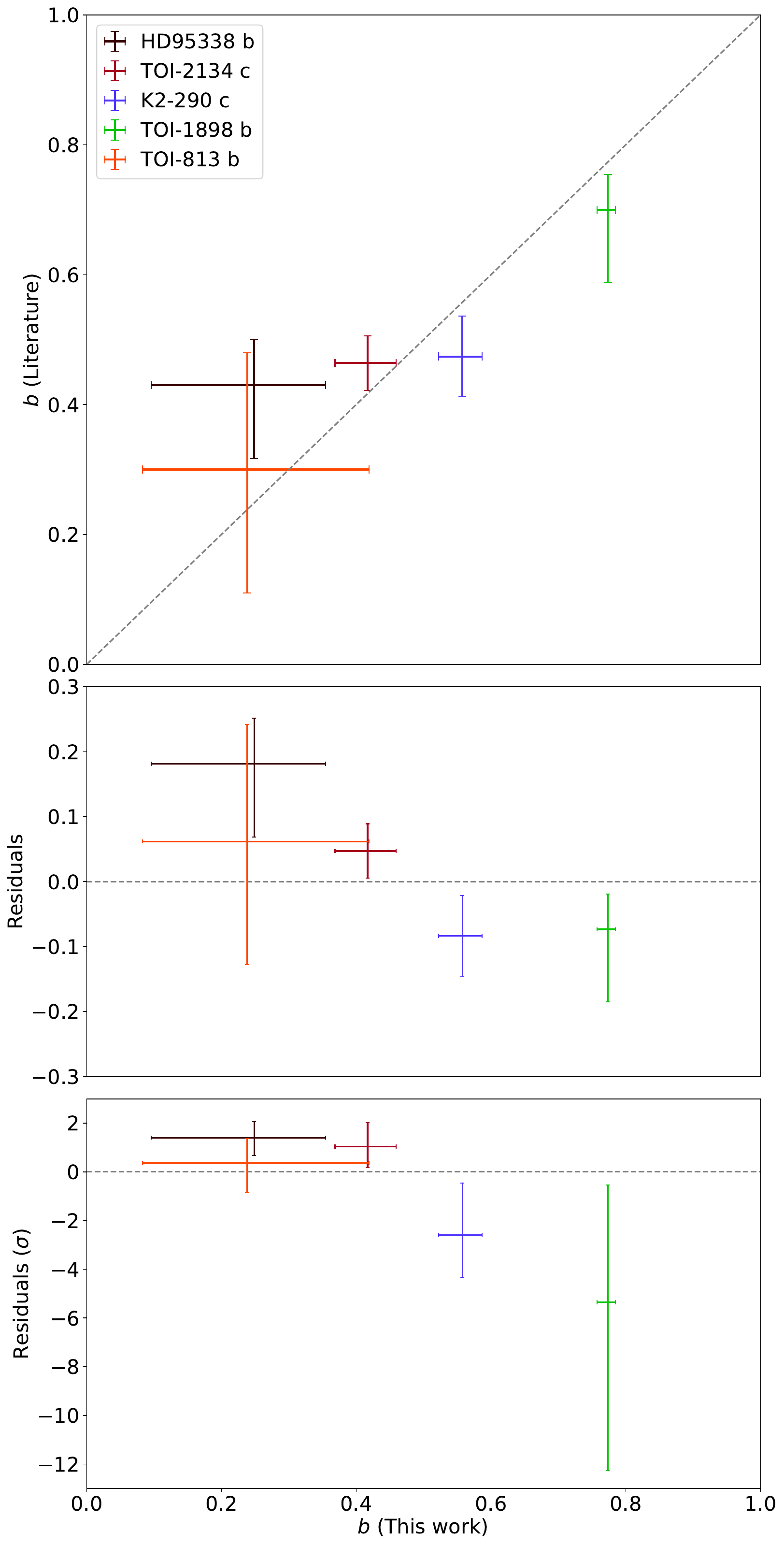}\vspace{0.2cm}
	\caption{Comparison plot of estimated values of $b$ from this work and the literature (top); the residuals from the comparison in parameter units (middle) and in uncertainties (bottom).}
	\label{fig:fig7}
\end{figure}

\begin{figure}
	\centering
	\includegraphics[width=\linewidth]{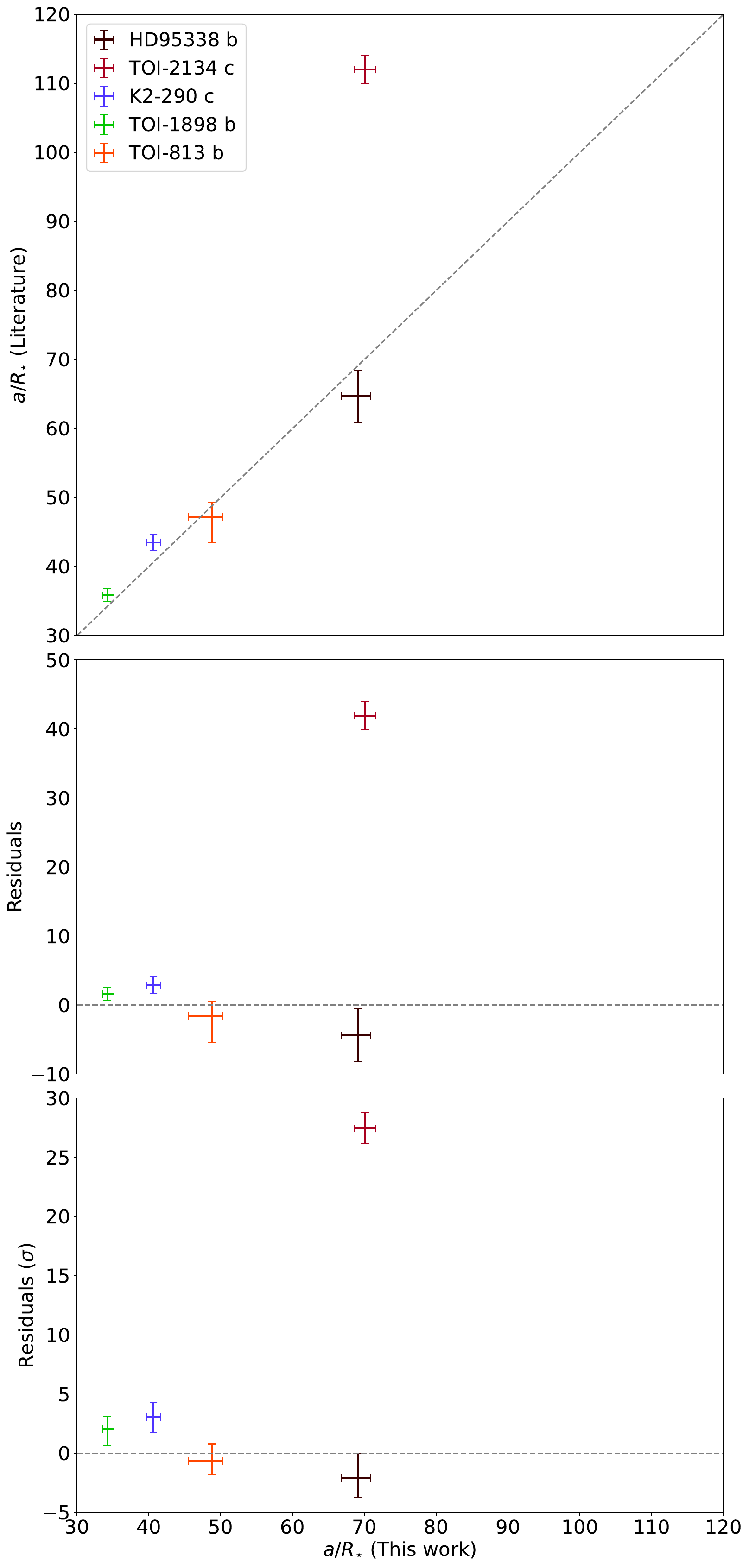}\vspace{0.2cm}
	\caption{Same as Figure \ref{fig:fig7}, but for $a/R_{\star}$.}
	\label{fig:fig8}
\end{figure}

\begin{figure}
	\centering
	\includegraphics[width=\linewidth]{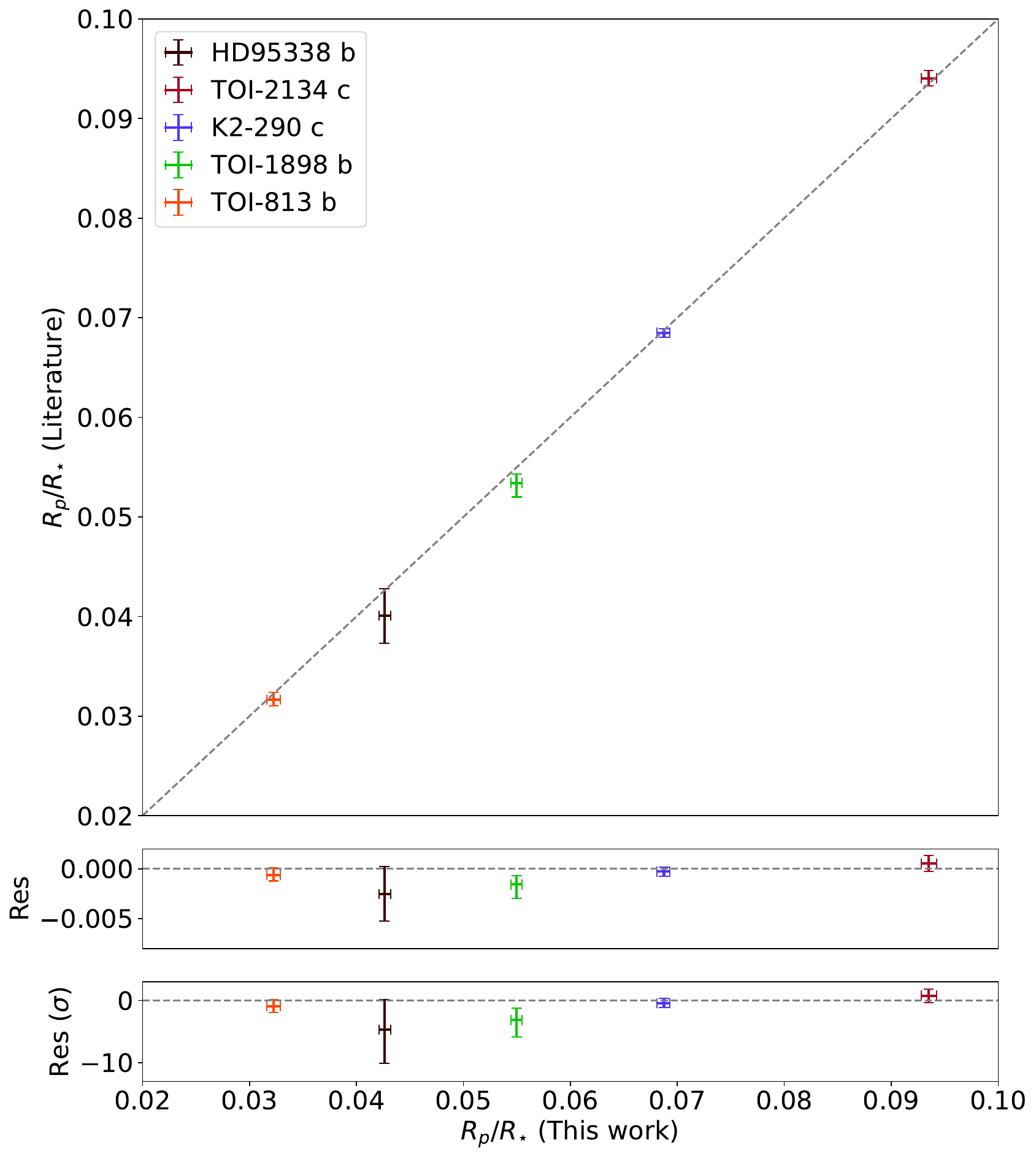}\vspace{0.2cm}
	\caption{Same as Figure \ref{fig:fig7}, but for $R_p/R_{\star}$. Only the values corresponding to the same telescopes were shown.}
	\label{fig:fig9}
\end{figure}

\begin{figure}
	\centering
	\includegraphics[width=\linewidth]{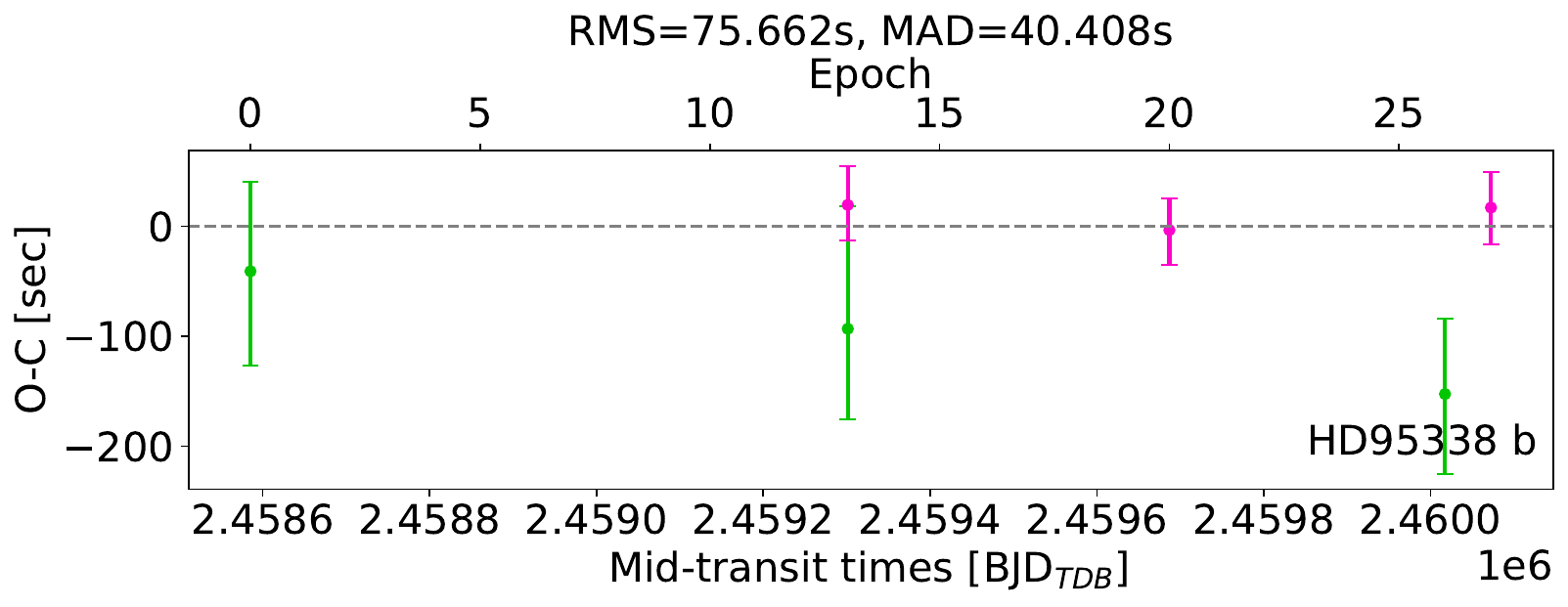}\vspace{0.3cm}
    \includegraphics[width=\linewidth]{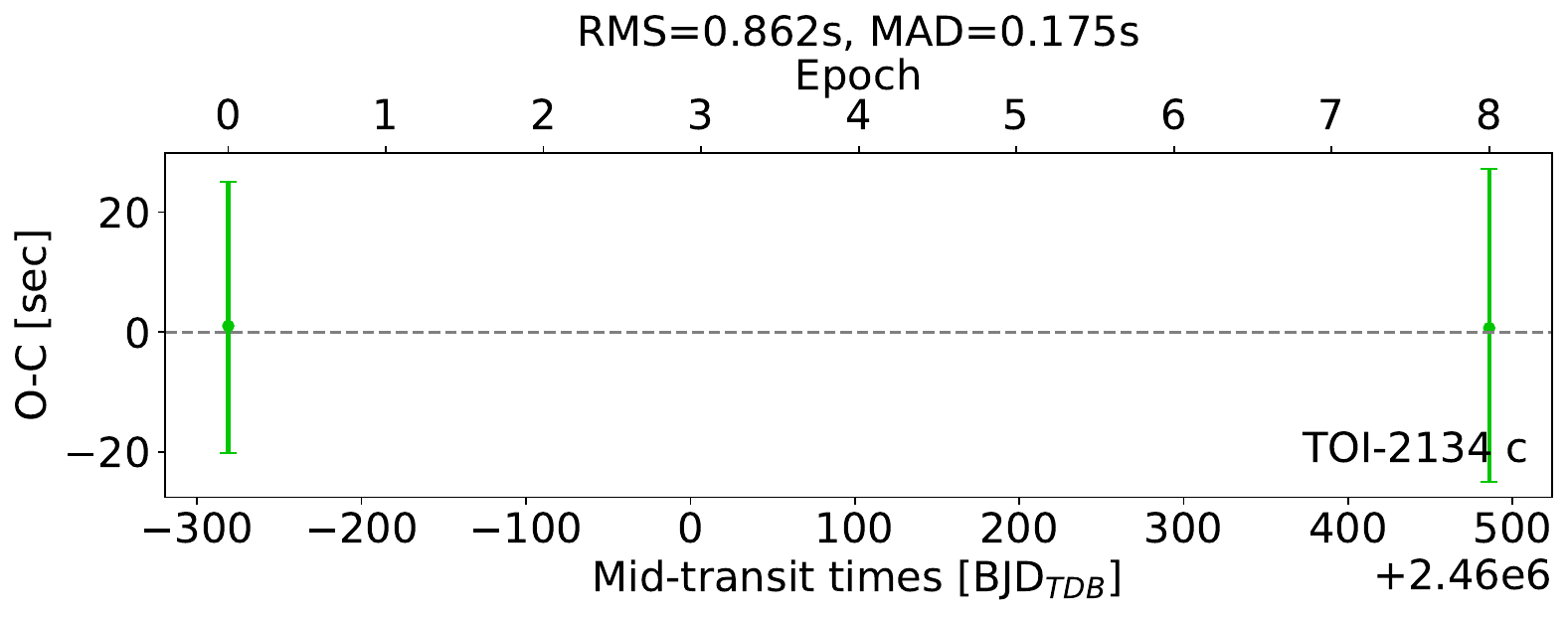}\vspace{0.3cm}
    \includegraphics[width=\linewidth]{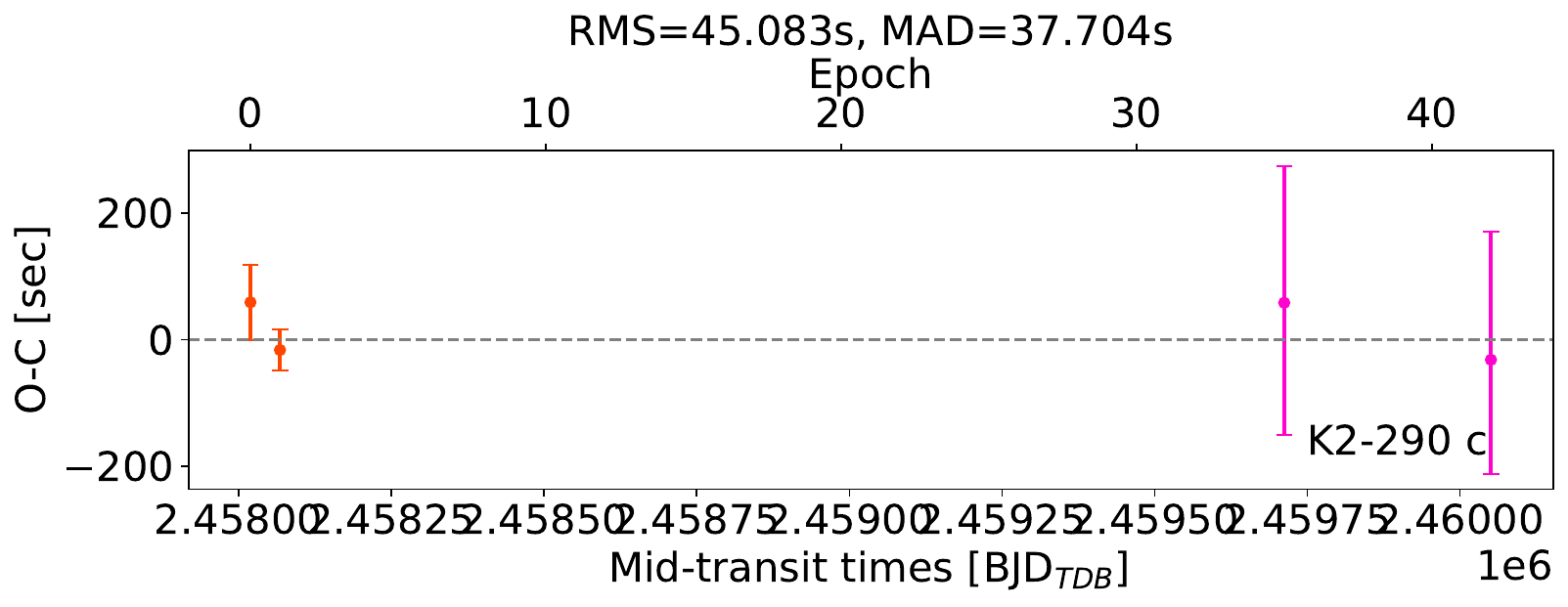}\vspace{0.3cm}
    \includegraphics[width=\linewidth]{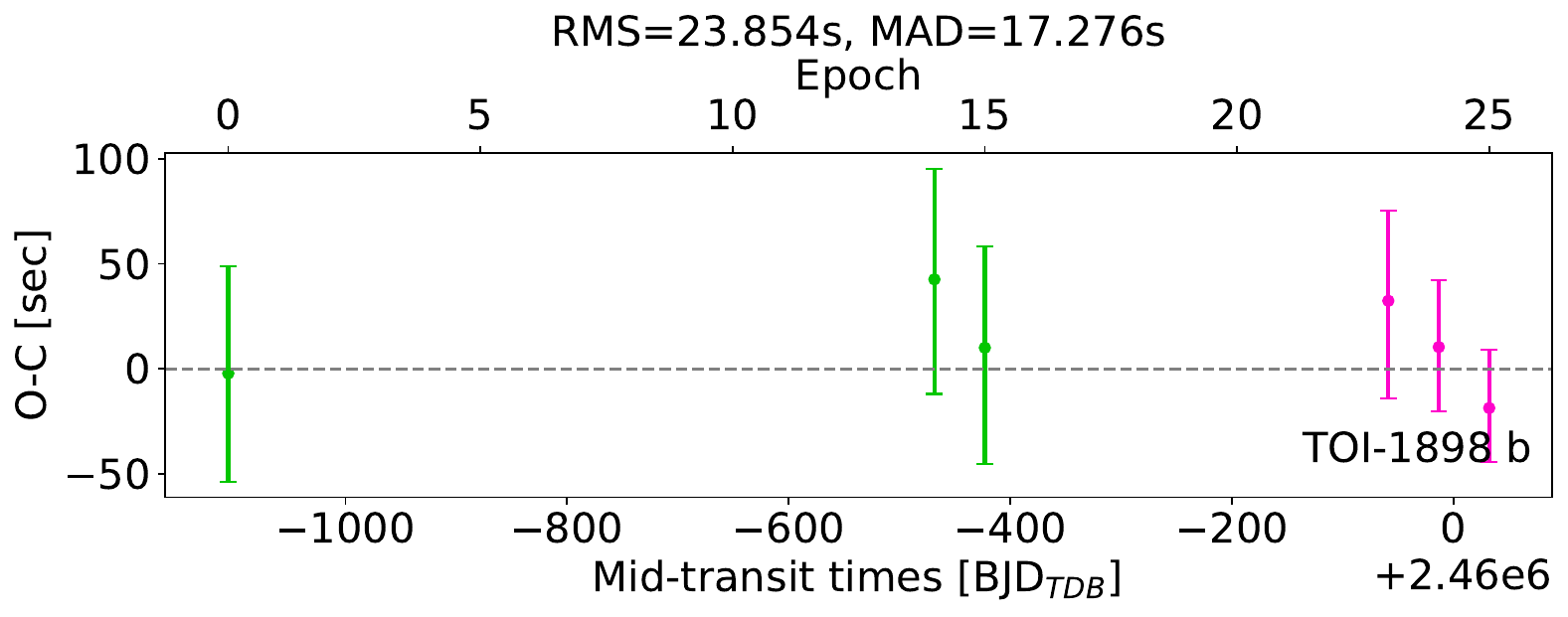}\vspace{0.3cm}
    \includegraphics[width=\linewidth]{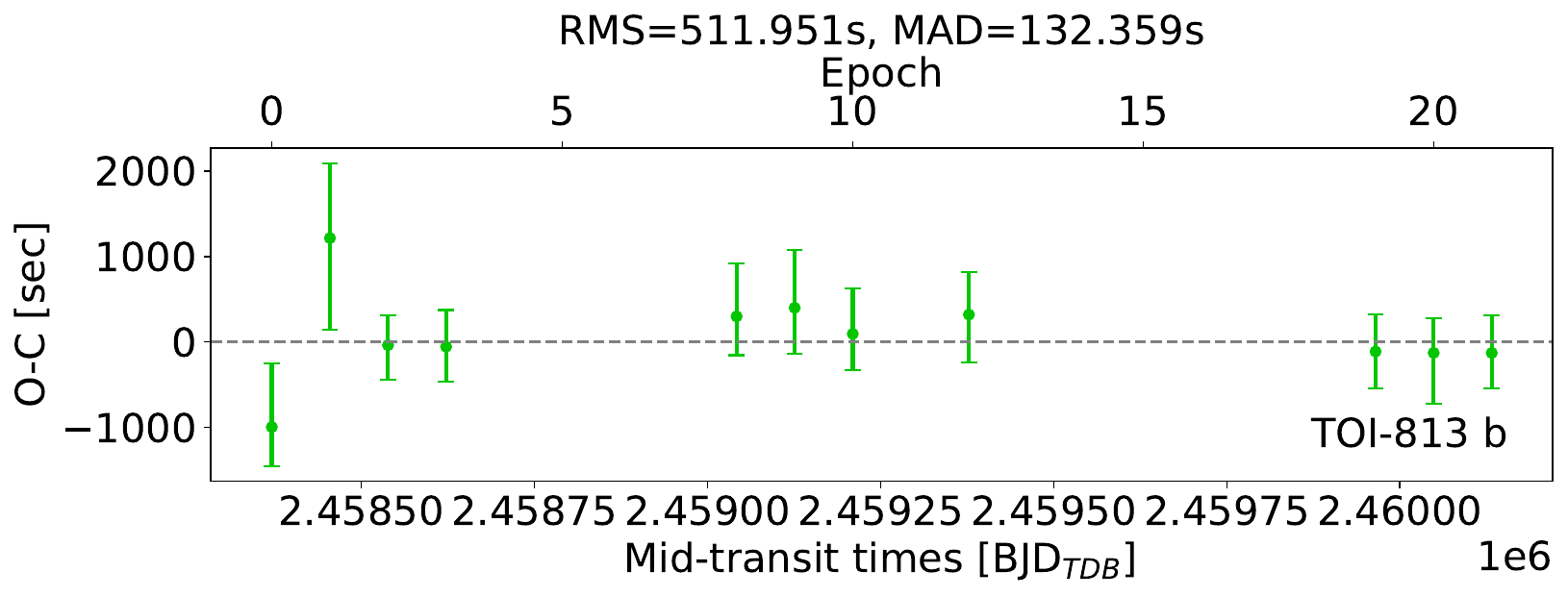}\vspace{0.2cm}
	\caption{O – C diagrams showing the estimated mid-transit times. The TESS, K2, and CHEOPS observations are shown in green, orange, and magenta respectively. The rms and median absolute deviation (MAD) of the mid-transit times have also been indicated.}
	\label{fig:fig6}
\end{figure}

\section{Discussions}\label{sec:sec4}

This study is aimed at updating the ephemeris and physical properties of a few high-value long-period transiting exoplanets, which would be immensely useful in future follow-up studies. The newly available observations for the five selected targets were analyzed in combination with the previously reported observations to obtain a better estimation of the parameters. The analyses incorporated a critical noise treatment algorithm to effectively reduce the impact of the noise components from various sources, thereby allowing for the best possible estimation of the parameters.

A comparison of the key estimated parameters with the literature values is shown in Table \ref{tab:tab6}. The present work has resulted in a more precise estimation of the ephemeris for all the targets, with the precision in the estimated periods being better than 5 seconds, except for TOI-813 b, for which the precision in the estimated period is better than 21 seconds. For HD95338 b, since only a single transit observation was reported in the previous study, the known ephemeris is improved significantly in this work, as five new transits were added. The uncertainty associated with the period is improved by $\sim$3 orders of magnitude, which will make it possible to calculate ephemeris precisely for future observations. Being among the long-period Neptunes with the brightest sources, HD95338 b is a very important target for future follow-up studies. Similarly, for TOI-2134 c, which also had only a single transit observation reported in the literature, the uncertainty associated with the period is improved by $\sim$3 orders of magnitude. Being a temperate zone gas-giant in a much shorter orbit than other similarly cold transiting planets, TOI-2134 c is a very interesting target for future follow-up studies, and thus the precise estimation of the ephemeris will be extremely useful.

For K2-290 c, the previous work consists of two subsequent transits from K2. Adding two new epochs after a considerable time scale has improved the uncertainty in the period estimation by approximately an order of magnitude. TOI-1898 b had three transits reported in the previous work, and the addition of three more transits has improved the uncertainties in the period by $\sim$3 times. For TOI-813 b, four transits were reported in the previous work. The addition of seven new transits has improved the uncertainties in the period estimation by more than an order of magnitude. All these improvements make these long-period exoplanets more readily accessible for future follow-up studies.

Apart from improvements in ephemeris, the addition of more transit lightcurves into analyses is also expected to improve the accuracy of the other estimated physical parameters. Since these parameters are widely adopted directly into the follow-up studies, a more accurate and verified set of parameters is beneficial, especially for these long-period exoplanets, where only a very limited volume of observations was used in the previous studies. Figures \ref{fig:fig7}-\ref{fig:fig9} depict the comparison of the parameters estimated from this work with the previous studies (from Table \ref{tab:tab6}). It can be noted from these figures that for most of the cases, the estimated parameters corrected the known values by statistically significant margins, and also improved the precision in the estimated parameters. Major notable improvements are seen in the estimated value of $b$ for TOI-1898 b, $a/R_\star$ for TOI-2134 c, and $R_p/R_\star$ for HD95338 b.

One of the major advantages of long-term follow-up studies of exoplanet transits is that any Transit Timing Variations (TTVs) over a long time scale can be probed. Especially, in the case of long-period exoplanets, which are not as strongly gravitationally bound to their host stars as their close-in counterparts, TTV signatures are expected to be more evident in the presence of other undetected planetary mass objects in these systems. Figure \ref{fig:fig6} shows the O-C diagrams of the estimated mid-transit times from this work. No significant TTV features have been detected for any of these targets, although small fluctuations could be visible in some cases. Significant longer-term follow-ups using existing facilities such as NGTS \citep{2018MNRAS.475.4476W} and CHEOPS \citep{2021ExA....51..109B}, as well as upcoming survey missions like PLATO \citep{2014ExA....38..249R} and ET \citep{2022arXiv220606693G} might be able to shed more light on the possibility of even longer-term variations in these targets.

\section{Conclusion}\label{sec:sec5}

The study involved a revisited analysis of transit photometric data of five very interesting long-period exoplanets, with an aim to provide better ephemeris and physical parameters for future follow-up studies. The analyses comprised observations previously reported as well as a significant number of new publicly available observations from TESS and CHEOPS. The data analyses and modeling were performed using a well-tested and effective critical noise treatment algorithm, which takes care of both correlated and uncorrelated noise in time.

The study has resulted in significant improvements in the ephemeris for all the targets, making them more accessible while planning future follow-ups. The study also resulted in significant improvements in the known physical parameters for these targets, which will be useful in the analyses of future follow-up programs. Leveraging the long-timescale of the available data, the presence of any TTV trend in these targets was surveyed, which has resulted in negative results. Combining this with the known TTVs in long-period exoplanet populations can be useful to understand the demographics of such long-period systems.

\section*{Acknowledgements}

I gratefully acknowledge the reviewer for their valuable feedback and insightful suggestions. I acknowledge Fondo Comité Mixto-ESO Chile ORP 025/2022 to support this research. The computations presented in this work were performed using the Geryon-3 supercomputing cluster, which was assembled and is maintained using funds provided by the ANID-BASAL Center FB210003, Center for Astrophysics and  Associated Technologies, CATA. This paper includes data collected by the TESS mission, which are publicly available from the Mikulski Archive for Space Telescopes (MAST). I acknowledge the use of public TOI Release data from pipelines at the TESS Science Office and at the TESS Science Processing Operations Center. Funding for the TESS mission is provided by NASA’s Science Mission directorate. Support for MAST is provided by the NASA Office of Space Science via grant NNX13AC07G and by other grants and contracts. This paper includes data collected by the Kepler mission and obtained from the MAST data archive at the Space Telescope Science Institute (STScI). Funding for the Kepler mission is provided by the NASA Science Mission Directorate. STScI is operated by the Association of Universities for Research in Astronomy, Inc., under NASA contract NAS 5–26555. This paper includes data collected by the CHEOPS mission. CHEOPS is an ESA mission in partnership with Switzerland with important contributions to the payload and the ground segment from Austria, Belgium, France, Germany, Hungary, Italy, Portugal, Spain, Sweden, and the United Kingdom. The CHEOPS Consortium would like to gratefully acknowledge the support received by all the agencies, offices, universities, and industries involved. Their flexibility and willingness to explore new approaches were essential to the success of this mission. CHEOPS data analyzed in this article will be made available in the CHEOPS mission archive (https://cheops.unige.ch/archive$\_$browser/).

\section*{Data Availability}

All the data used in this work are publicly available in the official archives as mentioned in Section \ref{sec:sec2}.

\bibliographystyle{mnras}
\bibliography{ms}

\end{document}